\documentclass[traditabstract]{aa}
\usepackage{graphicx}
\usepackage{amsmath,amssymb}
\usepackage{natbib}
\usepackage{enumerate}
\bibpunct{(}{)}{;}{a}{}{,}

\newcommand{\Msun}{M_\odot}

\newcommand{\Rsun}{R_\odot}
\newcommand{\MUSIC}{{\sf MUSIC}}

\begin{document}
   \title{Comparison of different nonlinear solvers for 2D time-implicit stellar hydrodynamics}
	\author{M. Viallet\inst{1,2}, I. Baraffe\inst{2}, R. Walder\inst{3}}

   \institute{
                Max-Planck-Institut f\"ur Astrophysik, Karl Schwarzschild Strasse 1, Garching, D-85741, Germany
                \and
                Physics and Astronomy, University of Exeter, Stocker Road, Exeter, UK EX4 4QL                
                \and
	      \'Ecole Normale Sup\'erieure, Lyon, CRAL (UMR CNRS 5574), Universit\'e de Lyon 1, France\\
             \email{mviallet@mpa-garching.mpg.de}
             }   
             
   \date{Received; accepted}

  \abstract{
Time-implicit schemes are attractive since they allow numerical time steps that are much larger than those permitted by the Courant-Friedrich-Lewy criterion characterizing time-explicit methods. This advantage comes, however, with a cost: the solution of a system of nonlinear equations is required at each time step. In this work, the nonlinear system results from the discretization of the hydrodynamical equations with the Crank-Nicholson scheme. We compare the cost of different methods, based on Newton-Raphson iterations, to solve this nonlinear system, and benchmark their performances against time-explicit schemes. Since our general scientific objective is to model stellar interiors, we use as test cases two realistic models for the convective envelope of a red giant and a young Sun. Focusing on 2D simulations, we show that the best performances are obtained with the quasi-Newton method proposed by Broyden. Another important concern is the accuracy of implicit calculations. Based on the study of an idealized problem, namely the advection of a single vortex by a uniform flow, we show that there are two aspects: i) the nonlinear solver has to be accurate enough to resolve the truncation error of the numerical discretization, and ii) the time step has be small enough to resolve the advection of eddies. We show that with these two conditions fulfilled, our implicit methods exhibit similar accuracy to time-explicit schemes, which have lower values for the time step and higher computational costs. Finally, we discuss in the conclusion the applicability of these methods to fully implicit 3D calculations.}

     
  \titlerunning{Comparison of different nonlinear solvers for 2D implicit hydrodynamics}
  \authorrunning{M. Viallet et al.}
   
  \maketitle

%
%
\section{Introduction}

The numerical integration of hydrodynamical equations in stellar interiors is characterized by numerical stiffness that can severely restrict the time step if a time-explicit scheme is used. 
Stiffness is due to sound waves traveling in deep stellar interiors that are usually characterized by low-Mach numbers. The flow is thus essentially frozen during the crossing time of a sound wave over a cell. Overcoming these short time scales imposed by sound waves may thus be of interest.  Radiative diffusion also causes stiffness, especially near the surface where diffusivity becomes strong. When using an explicit time-stepping method, numerical stiffness causes a restrictive stability condition on the time step. Therefore, depending on the physical case studied, one may choose to discretize advection and/or diffusion implicitly in order to relax the constraint on the time step. In this work, we consider the fully time-implicit discretization of the compressible hydrodynamical equations as implemented in \cite{viallet_towards_2011}, which provides a preliminary description of our time-implicit code  \MUSIC\footnote{``MUltidimensional Stellar Implicit Code"}.\

\MUSIC\ solves the equations describing the evolution of density, momentum, and internal energy, taking external gravity and radiative diffusion into account:

\begin{eqnarray}
\frac{\partial}{\partial t} \rho &=& - \vec \nabla \cdot (\rho \vec u),\\
\frac{\partial}{\partial t} \rho e &=& -\vec \nabla \cdot (\rho e \vec u) - p\vec \nabla \cdot \vec u + \vec \nabla \cdot (\chi \vec \nabla T),\\ 
\frac{\partial}{\partial t} \rho \vec u &=& - \vec \nabla \cdot (\rho \vec u\otimes \vec u)-\vec \nabla p + \rho \vec g,
\end{eqnarray}

\noindent where $\rho$ is the density, $e$ the specific internal energy, $\vec u$ the velocity, $p$ the gas pressure, $T$ the temperature, $\vec g$ the gravitational acceleration, and $\chi$ the thermal conductivity. For photons, the thermal conductivity is given by

\begin{equation}
\label{eq:chirad}
\chi = \frac{16 \sigma T^3}{3\kappa \rho},
\end{equation}

\noindent where $\kappa$ is the Rosseland mean opacity, and $\sigma$ the Stefan-Boltzmann constant. 

We follow the method of lines and perform the spatial discretization independently from the time discretization \citep[see e.g.][]{leveque2007finite}. The spatial discretization is based on finite volumes with staggered velocity components. The numerical fluxes are computed at interfaces with a monotonicity-preserving upwind method. We refer the reader to \cite{viallet_towards_2011} for details. After spatial discretization, we get a semi-discrete system

\begin{equation}
\label{eq:mol}
\frac{dU}{dt} = R(U),
\end{equation}

\noindent where $U = (\rho, \rho e, \rho \vec u)$, and $R$ contains the flux differencing and source terms. At this stage we introduce the implicit time-stepping method to discretize this system in time. In this work we  consider the second-order Crank-Nicholson method: 

\begin{equation}
U^{n+1} =  U^n + \frac{\Delta t}{2}\big ( R(U^{n+1}) + R(U^n) \big ).
\end{equation}

\noindent We define the nonlinear residual function

\begin{figure*}[t]
\parbox{0.5\linewidth}{\center \includegraphics[width=0.8\linewidth,clip=true,trim=20 20 20 20]{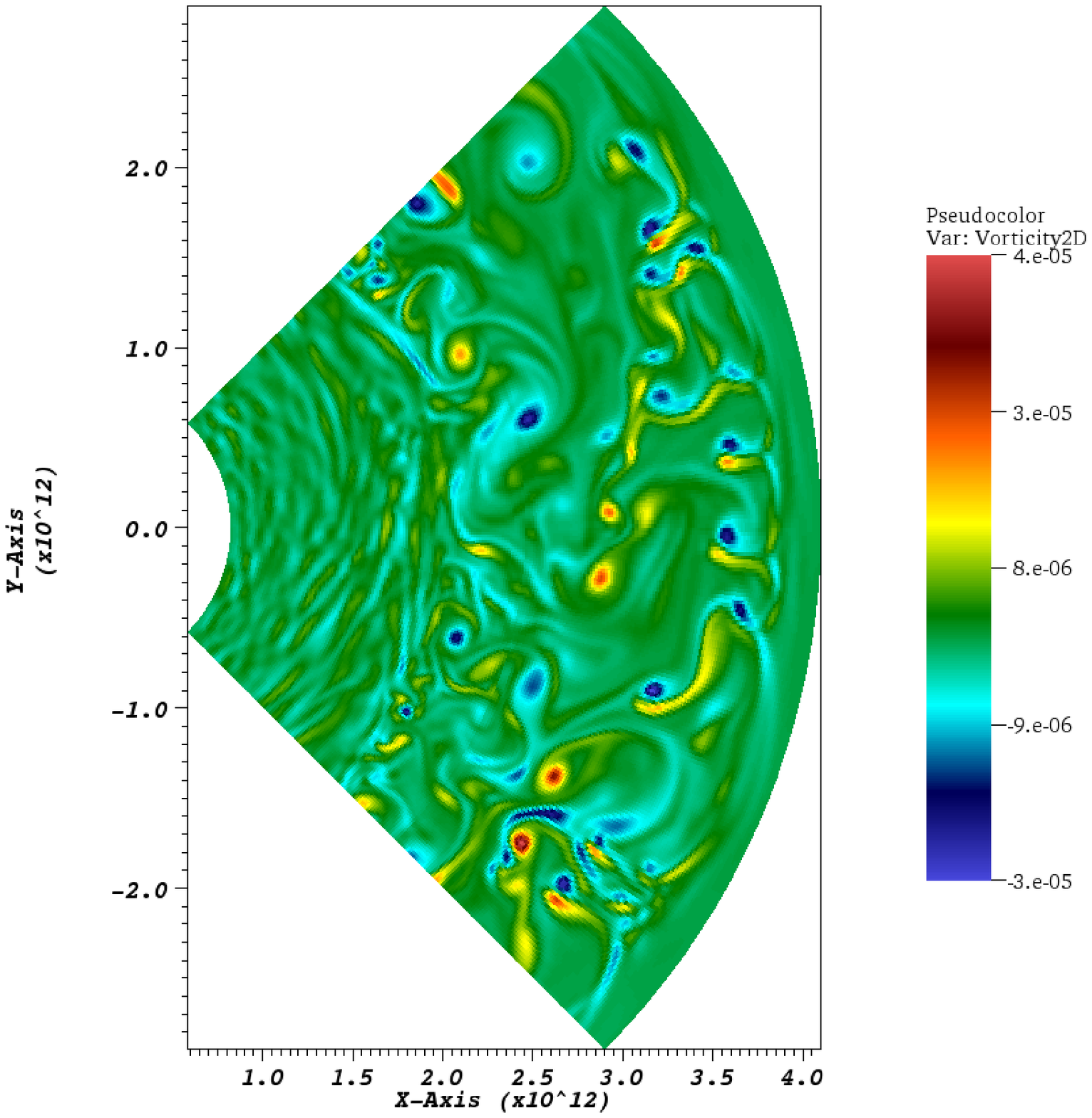}}
\parbox{0.5\linewidth}{\center \includegraphics[width=0.8\linewidth,clip=true,trim=20 20 20 20]{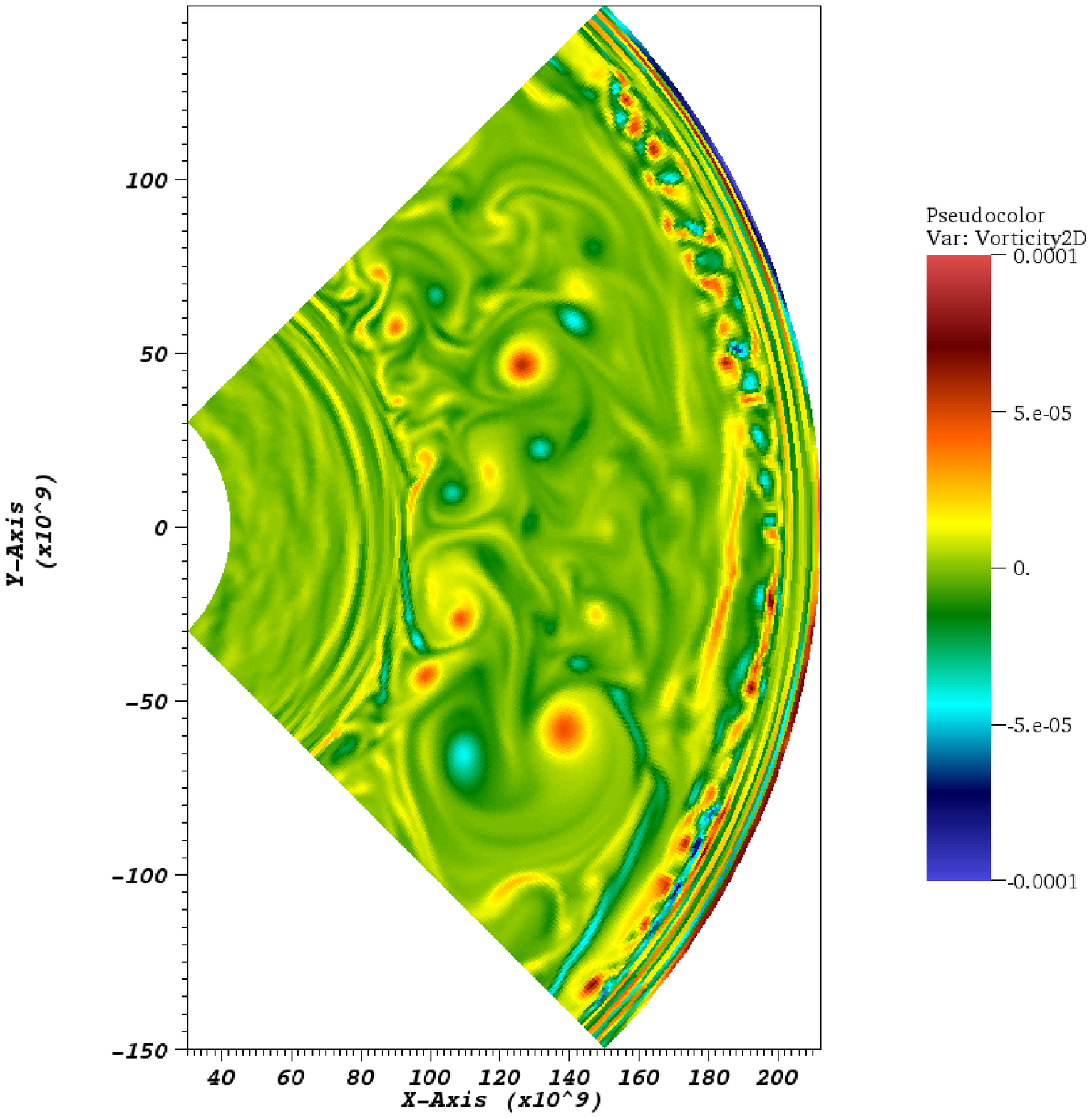}}
\caption{Initial models for the two test cases considered in this work. Left panel:  red giant model ($M=5 \Msun$, $R = 60 \Rsun$). Right panel: young Sun ($M=1 \Msun$, $R = 3 \Rsun$). Both panels show the vorticity field.}
\label{fig:initial_models}
\end{figure*}

\begin{equation}
\label{eq:crank_nicholson}
F(U, \Delta t) = \frac{U - U^n}{\Delta t} - \frac{1}{2}\big ( R(U) + R(U^n) \big ),
\end{equation}

\noindent so that $F(U^{n+1}, \Delta t)=0$ defines the new time step. For clarity, we drop the explicit dependence of $F$ on $\Delta t$ in the notation. Solving $F(U^{n+1})=0$, which allows the solution to evolve over time, is the challenge of implicit solvers. 
In this work, we present and compare different methods of solving this system of nonlinear equations. The nonlinear solvers considered here are all based on the Newton-Raphson method. Other approaches are possible, see for instance \cite{kifonidis_multigrid_2012} who use a multigrid method to solve the nonlinear system. The present work focuses on numerical methods to invert the nonlinear residual $F$ and does not address a comparison between different implicit time-stepping schemes (e.g. Crank-Nicholson, backward differentiation formulae, implicit Runge-Kutta methods). This is left for future work. Finally, we only consider 2D calculations, extension to 3D is currently under progress and is discussed in the conclusion.

The paper is organized as follows. In Sect.~\ref{testcases}, we present the test cases that we use to characterize the performances and accuracy of our nonlinear solvers. In Sect.~\ref{fullyimplicit}, we present in details the solution strategies for solving $F(U^{n+1})=0$. In Sect.~\ref{other_methods}, we present two other methods that are conditionally stable, namely the second-order Adam-Bashforth method and the minimum residual approximate implicit (hereafter MRAI) method. These methods are much cheaper, in terms of CPU time for a numerical time step, so we use them to benchmark the performances of the fully implicit methods. In Sect. \ref{results}, we present and discuss our results regarding the accuracy and the performances of the implicit solvers. In Sect. \ref{conclusion}, we conclude and outline future developments.

%
%
\section{Test cases}
\label{testcases}

We introduce two types of test cases. Since our scientific applications concern stellar hydrodynamics, we consider two realistic models of stellar convection to benchmark the performances of our nonlinear solvers. Flows in stellar interiors are turbulent, so they are no ideal ways to characterize the accuracy of the different methods, because pointwise convergence, for instance, is not meaningful due to the chaotic character of the flow. Therefore, to characterize the accuracy of the different nonlinear methods we consider an idealized problem: the advection of an isentropic vortex, as described in \cite{yee_entropy_2000}.

\subsection{2D stellar models}
\label{test_stellar_model}


Two different stellar models are used to benchmark the performance of our methods. The first one is the red giant model discussed in \cite{viallet_towards_2011}. The basic parameters of the stellar model are $M=5 \Msun$ and $R = 60 \Rsun$. The numerical setup is the same as in \cite{viallet_towards_2011}. The second test case corresponds to a young Sun with $M=1 \Msun$, $R = 3 \Rsun$, and an age of a few Myr, when a radiative core appears. The setup is thus  similar to that of the red giant test: the computational domain includes both the convective envelope and a fraction of the radiative core. As done for the red giant, a surface cooling term is introduced in the energy equation to mimic radiative cooling at the surface (see Viallet et al 2011). 

As in \cite{viallet_towards_2011}, we define three ``CFL" numbers:

\begin{align}
\mathrm{CFL}_\mathrm{hydro} &= \max \Big ( \frac{|u| + c_s}{\Delta x} \Big ) \Delta t, \\
\mathrm{CFL}_\mathrm{rad} &= \max \Big ( \frac{\chi}{\Delta x^2}  \Big ) \Delta t, \\
\mathrm{CFL}_\mathrm{adv} &= \max \Big ( \frac{|u|}{\Delta x}  \Big ) \Delta t \label{CFLadv},
\end{align}

\noindent where $c_s$ is the sound speed, $u$ the flow velocity, $\Delta x$ the typical mesh size, and $\Delta t$ the time step. These definitions are inspired by the well-known Courant-Friedrich-Lewy stability condition for explicit schemes. When choosing the value of the time step for numerical integration, $\mathrm{CFL}_\mathrm{hydro} \lesssim 1$ and $\mathrm{CFL}_\mathrm{rad} \lesssim 1$ usually provide very good guidelines for the maximum value that can be used if advection and/or radiative diffusion are solved explicitly. Note that $\mathrm{CFL}_\mathrm{adv} \lesssim 1$ is not a stability limit for compressible codes, but this number is introduced here as was recognized in \cite{viallet_towards_2011} as a useful quantity to monitor. $\mathrm{CFL}_\mathrm{adv} \gg 1$ implies that eddies are being advected over several cells during a time step, and this is expected to be prone to strong numerical damping. We investigate this in more detail in Sect. \ref{accuracy_dt}.

Both test cases use the same equation-of-state and opacity routines, and are discretized with the same resolution of $216 \times 256$. From the raw numerical cost, the two cases are therefore identical. However, they differ in terms of the typical convective velocity: in the red giant model the convective Mach number is on the order of 0.1, whereas in the young Sun model it is on the order of 0.01. On average, a time step that corresponds to $\mathrm{CFL}_\mathrm{adv} \sim 1$ translates into $\mathrm{CFL}_\mathrm{hydro} \sim 40$ ($\Delta t \sim$ 0.15 d) in the red giant model and into $\mathrm{CFL}_\mathrm{hydro} \sim 500$ ($\Delta t  \sim$ 1.5 h) in the young Sun model. These test cases are therefore useful for investigating how the performances of the different methods change for different CFL regimes. Finally, radiative diffusion is not a very stiff process in these models. Radiative diffusion becomes stiff close to the surface layers, where the density decreases significantly \citep[see e.g. Fig. 13 in][]{viallet_towards_2011}. These surface layers are not described realistically in our simulations, because they are modeled by an artificial isothermal region. Therefore, in the models discussed here, numerical stiffness stems essentially from the sound waves, and our tests address the efficiency of discretizing advection implicitly.


%

\subsection{Advection of an isentropic vortex in 2D}
\label{test_advection}

This test problem was originally described in \cite{yee_entropy_2000}. Here we closely follow the setup and parameters of \cite{kifonidis_multigrid_2012}\footnote{There is a sign error in their Eq. (C.2). Our equation (\ref{eq:dt}) is correct, see also \cite{yee_entropy_2000}. }. The initial state consists of an isentropic vortex (i.e. zero entropy perturbation) embedded in an uniform flow characterized by $u_\infty=1,\ v_\infty=0,\ \rho_\infty=1,\ T_\infty=1$. The vortex corresponds to the following perturbations in the state variables:

\begin{align}
(\delta u, \delta v) &= \frac{\hat{\beta}}{2\pi} \mathrm{e}^{\frac{1-r^2}{2}} (-\bar{y}, \bar{x}),\\
\delta T &= - \frac{(\gamma-1) \hat{\beta}^2}{8 \gamma \pi^2} \mathrm{e}^{1-r^2}, \label{eq:dt}
\end{align}

\noindent where $T=p/\rho$, $\gamma$ is the adiabatic index, and $\hat{\beta}$ the vortex strength. We use here $\gamma=1.4$ and $\hat{\beta}=0.75$.

The initial conditions are

\begin{align}
\rho &= (T_\infty + \delta T)^{\frac{1}{\gamma-1}} \\
u &= u_\infty + \delta u \\
v &= v_\infty + \delta v \\
e &= \frac{\rho^{\gamma-1}}{\gamma-1}.
\end{align}

The computations are performed on a 2D Cartesian domain $[-4,4]\times[-4,4]$. Initially,  the vortex is centered on the origin. The advection of the vortex is computed numerically during 0.4 unit of time. The exact solution of the problem corresponds to the vortex profile being shifted by 0.4 unit of length in the $x$ direction. In Sect. \ref{accuracy}, we characterize the accuracy of our nonlinear methods by comparing the computed density field $\rho_{i,j}$ and the expected analytical solution $\rho^0_{i,j}$, using three different norms:

\begin{align}
& L_1\mathrm{-error:\ } || \rho - \rho^0 ||_1 = \frac{1}{N_x N_y} \sum_{i,j} | \rho_{i,j} - \rho^0_{i,j} |, \\
& L_2\mathrm{-error:\ } || \rho - \rho^0 ||_2 = \sqrt{ \frac{1}{N_x N_y} \sum_{i,j} ( \rho_{i,j} -\rho^0_{i,j} )^2 }, \\
& L_\infty \mathrm{-error:\ } || \rho - \rho^0 ||_\infty = \max_{i,j} | \rho_{i,j} - \rho^0_{i,j} |,
\end{align}

\noindent $N_x$, $N_y$ being the grid dimensions.
%
%
\section{Implicit method: nonlinear solvers}
\label{fullyimplicit}

\begin{table}[t]
   \caption{Summary of the Newton-Raphson method.}
   {\bf Goal} - Solve $F(U^{n+1}) = 0$\\
   {\bf Input} - $U^{n}$: solution at time $t_n$. Nonlinear residual $F(U)$. Time step $\Delta t$.\\
   {\bf Output} - $U^{n+1}$: solution at time $t_{n+1} = t_n + \Delta t$. 
\begin{enumerate}
	\item Set the initial guess $U^{(0)}=U^{n}$.
	\item At iteration $k$:
		\begin{enumerate}[i.]
		\item Solve a linear system of the form $J^{(k)} \delta U^{(k)} = - F(U^{(k)})$ to get the Newton direction $\delta U^{(k)}$;
		\item Compute $U^{(k+1)} = U^{(k)} + \delta U^{(k)}$;
		\item If max $|| \delta U^{(k)} / U^{(k)} || < \epsilon$, go to 3;
		\end{enumerate}
	\item Adapt $\Delta t$ based on the time step strategy (see text);
	\item Set $U^{(n+1)} = U^{(k)}$.
\end{enumerate}   
   \label{tab:generalNR}
\end{table}

\subsection{Classes of solvers}
\label{solver_classes}

The general framework for solving $F(U^{n+1})=0$ is the well-known Newton-Raphson method, as summarized in Table \ref{tab:generalNR}. Starting with an initial guess, here the solution at time step $n$, one starts an iterative process. At each iteration, a linear system is solved to get the Newton correction $\delta U^{(k)}$ (see step 2. ii. in Table \ref{tab:generalNR}), which is used to update the current iteration. Convergence is tested on the relative corrections, which are required to become smaller than a given ``nonlinear tolerance" $\epsilon$. The first iteration to fulfill this condition is then the new solution $U^{n+1}$. From this general framework, we can distinguish different classes of solvers, described below. 

\begin{figure}[t] 
   \centering
   \includegraphics[width=0.9\linewidth]{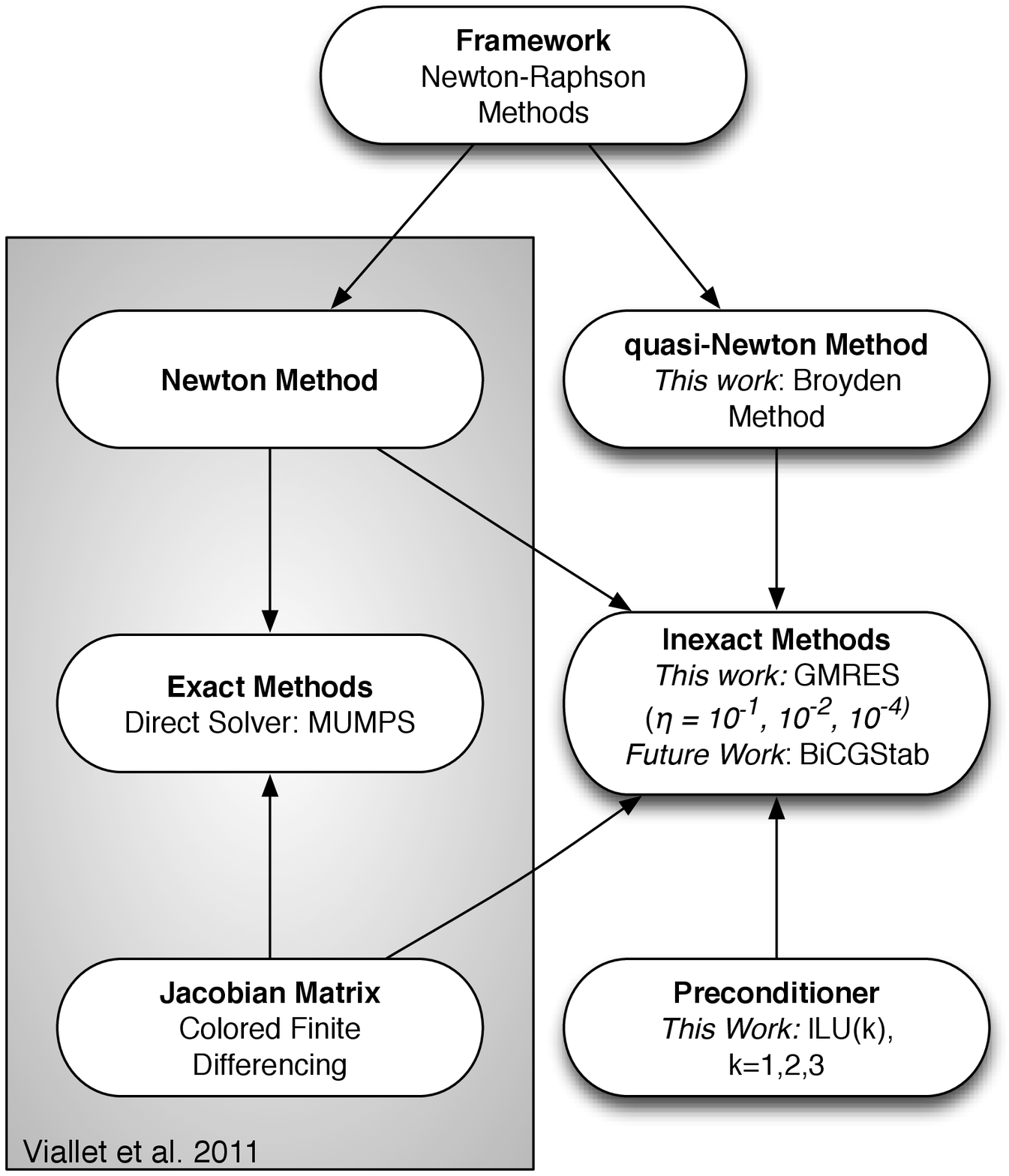} 
   \caption{Summary of the different classes of Newton-Raphson solvers. The shaded area refers to the approach described in \cite{viallet_towards_2011}.}
   \label{fig:summary}
\end{figure}

One can first distinguish  ``direct"  from ``iterative" methods of solving the linear system. A direct method solves the system in a finite number of steps, using e.g. Gaussian elimination. Without rounding errors, it would deliver the exact solution, therefore these methods are often referred to as ``exact''. In practice, however, the numerical accuracy depends on the condition number of the matrix, the concrete exact method, and its numerical implementation.  In particular for large systems, the accuracy can become very poor even if elaborate pivoting is used. On the other hand, an iterative method produces a sequence of approximations that converge to the solution and allows the user to tune the accuracy of the result. The motivation for these methods is the counter-productivity of ``over-solving'' the linear system within Newton-Raphson iterations, given that what is essentially required is an update direction that lowers the norm of the nonlinear residual. Approximate solutions of the linear system can fulfill this requirement and, even if the number of Newton-Raphson iterations increases, an overall gain in performance is observed. For this reason, this type of method is usually referred to as ``inexact". 

The most efficient iterative methods for solving sparse linear systems seek a solution in the so-called Krylov space of the matrix, see e.g. Chapter 6 in \cite{Saad:2003:IMS:829576}. For our purpose, an important requirement for these ``Krylov solvers" is that they can deal with general nonsymmetric matrices. Here, we use the generalized minimum residual method, hereafter GMRES, described in \cite{Saad:1986:GGM:14063.14074}. We use the ``restarted" variant of GMRES, i.e. we limit the dimension of the Krylov space to 40 and the method is restarted from the last solution when the number of iterations reaches this number. 

Another popular method, which we plan to test in the future, is the stabilized biconjugate gradient method \citep[BiCGStab, see][]{vanderVorst:1992:BFS:131916.131930}. The combination of a Krylov solver with Newton-Raphson iterations is often called a ``Newton-Krylov'' method. We define the forcing term, or ``linear tolerance", $\eta$, as the parameter that controls the accuracy of the iterative solver, based on the following criterion for convergence:

\begin{equation}
\label{eq:forcing_term}
|| J^{(k)} \delta U^{(k)} + F(U^{(k)})||_2 < \eta ||F(U^{(k)})||_2.
\end{equation}

We consider linear tolerances in the range $10^{-1} - 10^{-6}$. Finally, iterative methods need ``preconditioning'' to be efficient, or even to achieve convergence. This is discussed in Sect. \ref{precond}.


One can also distinguish between ``Newton''  and ``quasi-Newton'' methods.  In a Newton method, the matrix $J^{(k)}$  (see Table 1) is the Jacobian matrix evaluated at $U^{(k)}$:

\begin{equation}
J^{(k)} = \frac{\partial F}{\partial U} (U^{(k)}),
\end{equation}

\noindent whereas in a quasi-Newton method, $J^{(k)}$ is only an approximation of the Jacobian matrix. Newton methods are characterized by a quadratic convergence near the solution, which is not the case for quasi-Newton methods. However, the latter usually performs better (and are designed to), because the quasi-Jacobian matrix is typically cheaper to compute than the Jacobian matrix. An often suggested choice for such a ``quasi-Jacobian'' matrix is the Jacobian of a simplified system of equations, resulting for instance from a lower-order spatial discretization. Here, we use the method proposed by \cite{broyden_1965}. The Broyden method is a generalization of the secant method, with $J^{(k)}$, the so-called ``Broyden" matrix, initialized with the Jacobian matrix at the first Newton-Raphson iteration. It is then updated during the next nonlinear iterations using the formula proposed by \cite{broyden_1965}:

\begin{equation}
\label{eq:broyden_update}
J^{(k+1)} = J^{(k)} + \frac{\big(\delta F^{(k)} - J^{(k)}\delta U^{(k)} \big) (\delta U^{(k)})^T }{|| \delta U^{(k)}||_2},
\end{equation}

\noindent where $\delta F^{(k)} = F(U^{(k+1)}) - F(U^{(k)})$, and $J^{(k)}$ is the previous Broyden matrix. It fulfills the following relation:

\begin{equation}
J^{(k+1)}  \big ( U^{(k+1)} - U^{(k)} \big ) = F(U^{(k+1)}) - F(U^{(k)}),
\end{equation}

\noindent which can indeed be interpreted as an approximation of the Jacobian matrix using the secant method.

%

\cite{viallet_towards_2011} present results obtained with an exact Newton method. The use of a state-of-the-art direct solver, {\sf MUMPS}\footnote{see http://graal.ens-lyon.fr/MUMPS/} \citep[see][]{amestoy_fully_2001,amestoy_hybrid_2006}, resulted in a robust, but expensive method that is clearly outperformed by the methods presented here.


\MUSIC\ is interfaced with the {\sf Trilinos} toolkit \cite[see][]{heroux_overview_2005}, which provides access to state-of-the-art libraries. We use the library {\sf Epetra} for matrix/vector storage and basic linear algebra operations, {\sf AztecOO} for Krylov solvers, {\sf IFPACK} for preconditioners, and {\sf NOX} for computing the Jacobian matrix and implementing the Broyden method.

Figure \ref{fig:summary} summarizes the different strategies for solving the nonlinear system resulting from an implicit discretization. This work focuses on inexact methods. The next two sections describe the method for computing the Jacobian matrix and the preconditioning strategy for the GMRES solver, focusing on the stellar models described in Sect. \ref{test_stellar_model}.

\begin{table*}[t]
\caption{Costs for computing the Jacobian matrix, ILU preconditioners, and for performing 20 GMRES iterations with preconditioning operation. Tests are performed on the stellar models. Times are given in seconds. Single core computations (CPU: Intel Xeon Westmere at 2.80 Ghz).}
\centering
\begin{tabular}{l c c c c c c c c}
\hline \hline
Resolution & Jacobian & ILU(1) & ILU(2) & ILU(3) & 20 GMRES & 20 GMRES & 20 GMRES & 20 GMRES\\
 &  & & & & w/o preconditioner & w/ ILU(1) & w/ ILU(2) & w/ ILU(3)\\
\hline
$216\times256$ & 7.46 & 2.03 & 4.74 & 10.06 & 0.44 & 1.07 & 1.48 & 2.23 \\
$432\times512$ & 31.69 & 8.29 & 19.27 & 40.68 & 2.25 & 4.72 & 6.39 & 9.63 \\
$872\times1024$ & 134.61 & 53.25 & 81.10 & 205.85 & 8.42 & 26.31 & 27.51 & 41.07 \\
\hline
\end{tabular}
   \label{tab:costs}
\end{table*}

\begin{figure}[t]
   \centering
   \includegraphics[width=0.9\linewidth]{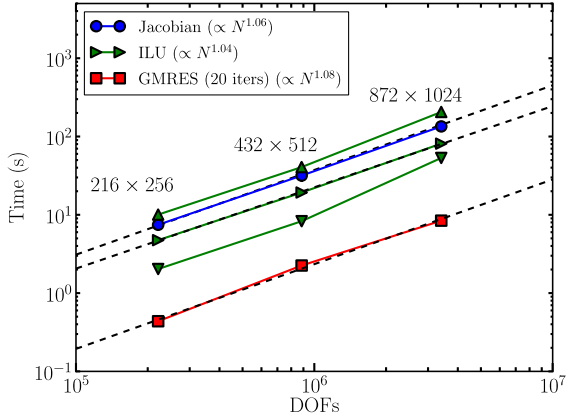}
   \caption{Cost of the linear algebra elementary operations versus the number of degrees of freedom (DOF): computation of the Jacobian matrix, construction of the ILU preconditioners (down-triangles: ILU(1), right-triangles: ILU(2), up-triangles: ILU(3)), and 20 GMRES iterations \emph{without preconditioning}. Tests are performed on the stellar models. Single core computations (CPU: Intel Xeon Westmere at 2.80 Ghz).}
   \label{fig:linear_algebra_cost}
\end{figure}

\subsection{Jacobian matrix computation}
\label{jacobian}

We compute the Jacobian matrix elements by finite differencing:

\begin{equation}
J^{(k)}(U) \approx \frac{F(U^{(k)} + \delta U) - F(U^{(k)})}{\delta U}.
\end{equation}

\noindent This strategy is chosen because it provides flexibility to change the code, in terms of both implemented physics (e.g. source terms, physical equations) and numerical method (e.g. limiters, reconstruction order). We use a forward formula to compute the Jacobian elements. A centered formula is two times more expensive, and it is not clear that the improved accuracy would significantly affect the nonlinear convergence, especially within a quasi-Newton strategy. As in \cite{viallet_towards_2011}, we use the colored finite differencing algorithm \citep[hereafter CFD, see][]{curtis_estimation_1974,gebremedhin_what_2005}. Since the main cost is in evaluating $F$, CFD minimizes the number of function evaluations by grouping independent columns of the Jacobian in a ``compressed" representation of the matrix. Technically, this is done prior to the computation using the sparsity graph of the Jacobian, which is known in advance since it only depends on the physical equations and numerical method. In \cite{viallet_towards_2011}, the number of columns of this compressed representation, also called the number of colors $n_g$, was found to be roughly $50$ for the discretization of the hydrodynamical equations in 2D. The strategy is then to loop on colors, to perturb all variables of that color, to recompute $F$, and to apply finite differencing to obtain several matrix coefficients at once. In total, the CFD algorithm needs $n_g+1$ evaluations of $F$ to compute the Jacobian matrix. The advantage of the CFD method is that $n_g$ is roughly independent of the problem size so that the cost of the algorithm scales linearly with the matrix size. Figure \ref{fig:linear_algebra_cost} (see also Table \ref{tab:costs}) illustrates the cost of the method for our stellar models at three different resolutions: $216\times256$, $436\times512$, and $872\times1024$. 

Finally, another possibility for computing the Jacobian matrix is to use ``automatic differentiation", see e.g. \cite{Griewank2008EDP}. The goal of automatic differentiation is to compute the Jacobian matrix by achieving the performances, both in time and in accuracy, of an analytically hand-coded Jacobian. We plan to test this in the future.

\subsection{ILU preconditioning}
\label{precond}

\begin{figure*}[t] 
   \centering
   \parbox{0.49\linewidth}{\includegraphics[width=0.9\linewidth]{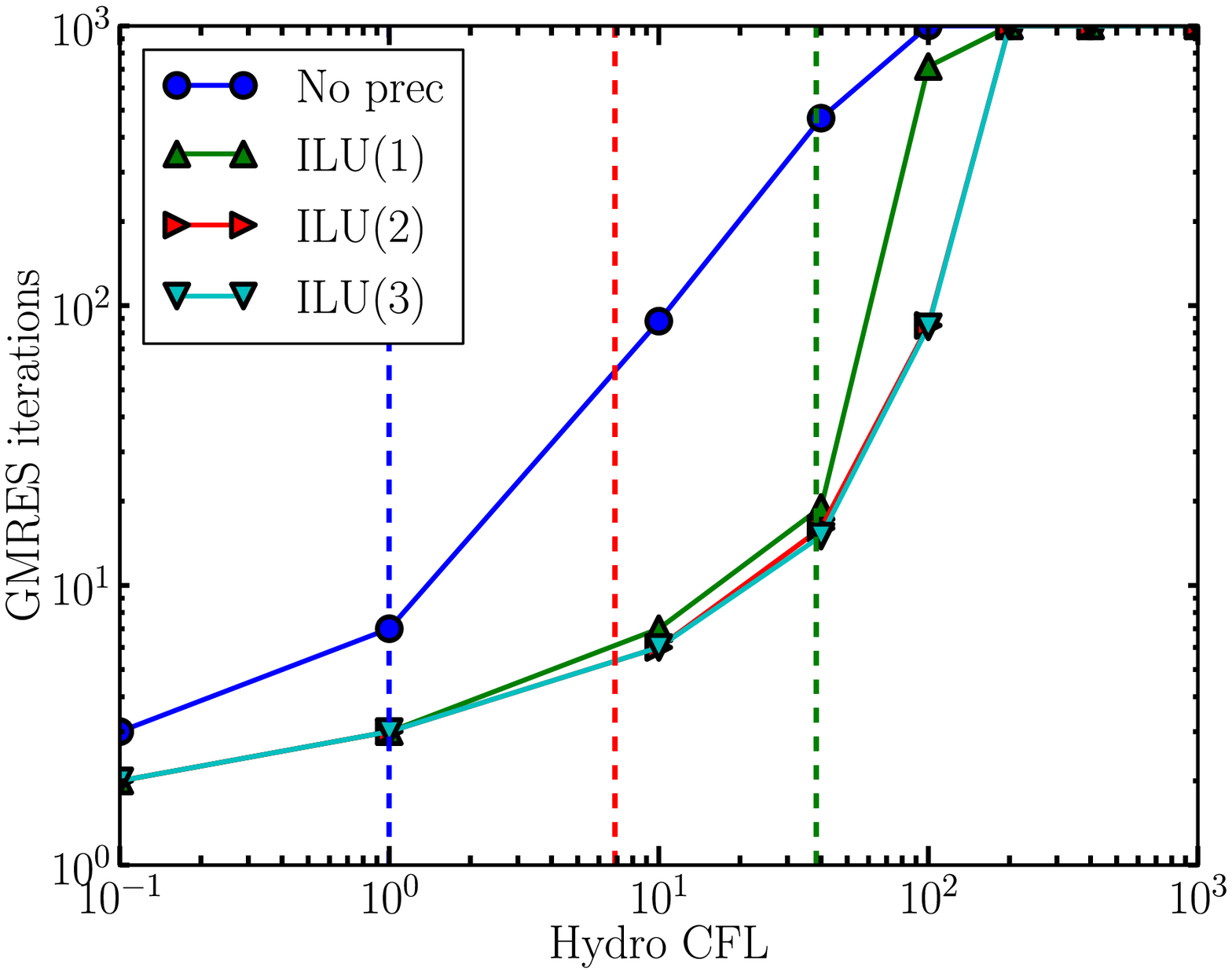}}
   \parbox{0.49\linewidth}{\includegraphics[width=0.9\linewidth]{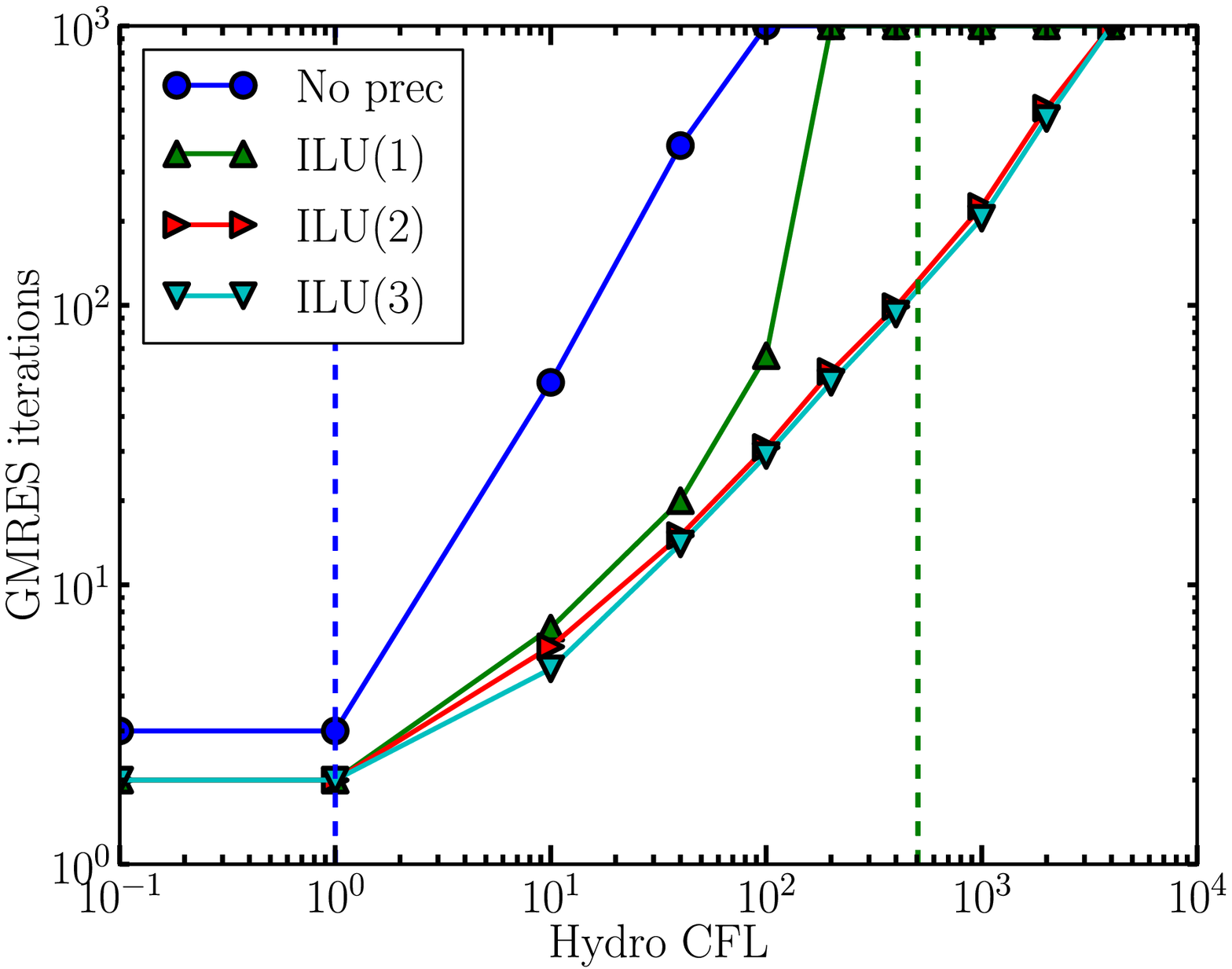}}
   \caption{Number of GMRES iterations for different CFL$_\mathrm{hydro}$ and different preconditioning strategy (including no preconditioning) for the red giant (left panel) and the young Sun (right panel) models. The tolerance of the linear solver is $\eta = 10^{-4}$ and the maximum number of iterations is set to 1000. The vertical dashed lines denote the position of CFL$_\mathrm{hydro}=1$ (blue line), CFL$_\mathrm{adv}=1$ (green line), CFL$_\mathrm{rad}=1$ (red line). In the right panel, CFL$_\mathrm{rad}=1$ falls outside of the figure.}
   \label{fig:preconditioner_giant}
\end{figure*}

In Fig. \ref{fig:preconditioner_giant}, the blue dotted lines illustrate how the number of GMRES iterations needed to achieve convergence with $\eta=10^{-4}$ increases with the time step, measured here in terms of CFL$_\mathrm{hydro}$, in both the red giant and young Sun models. The iterative solver eventually fails at converging within the allowed number of iterations, set here to the rather high value of 1000. The bad performance of the iterative method in this case is due to the condition number of the Jacobian matrix, which increases with the time step. This can be qualitatively understood by the following argument. The Jacobian matrix $J$ corresponding to Eq. (\ref{eq:crank_nicholson}) can be written as

\begin{equation}
J(U) = \frac{\mathbb{I}}{\Delta t} - \frac{1}{2} \frac{\partial R}{\partial U}(U).
\end{equation}

\noindent For low values of the time step, $J \approx \frac{\mathbb{I}}{\Delta t}$ and it has spectral properties that are very similar to the unity matrix, although eigenvalues are here clustered around $1/\Delta t$. As a result the matrix is diagonally dominant and its condition number is close to one, so that the iterative method performs very well. On the other hand, for high values of the time step, $J \approx -\frac{1}{2}\frac{\partial R}{\partial U}$ which is much more complex and has less favorable spectral properties. For instance, $\frac{\partial R}{\partial U}$ has high off-diagonal values that deteriorate the condition number of the matrix. If we denote by $\lambda$ the eigenvalues of $\frac{\partial R}{\partial U}$, the eigenvalues of $J$ are $1/\Delta t - \frac{\lambda}{2}$, and the unity matrix will dominate the spectral properties of $J$ when

\begin{equation}
\Delta t | \lambda | \lesssim 2,
\end{equation}

\noindent for all $\lambda$. For a hyperbolic equation discretized with an upwind method, the eigenvalues $\lambda$ are located in the left-hand region of the complex plane, and the maximum value of the module is close to the velocity of the characteristics (e.g. sound waves) divided by the mesh size $\Delta x$ \citep[see][Sect. 10.4]{leveque2007finite}. As a consequence, the above inequality is similar to the CFL condition, and indeed Fig. \ref{fig:preconditioner_giant} shows that the number of iterations significantly increases as soon  as CFL$_\mathrm{hydro}$ becomes larger than one. Since implicit methods target CFL$_\mathrm{hydro} \gg 1$, one will always face this problem. However, to obtain good performances it is important to keep the number of GMRES iterations to a moderate value (a few dozen). To achieve that, one has to use preconditioning. The preconditioned GMRES method solves the following modified system

\begin{equation}
\label{eq:lin_sys_rp}
J M^{-1} \big( M \delta U \big) = - F(U),
\end{equation}

\noindent where $M$ is the preconditioning matrix. We dropped the Newton-Raphson index $(k)$ for clarity. If $M$ is a good enough approximation of $J^{-1}$, the number of iterations needed for convergence will be much lower. It is also important to consider the cost for computing the preconditioning matrix and for the preconditioning operation during GMRES iterations. There is therefore a trade-off between the quality of $M$, in terms of an approximation of $J^{-1}$, and the cost involved. 

In this work, we use an incomplete LU (ILU) factorization of the Jacobian matrix as a preconditioner, see e.g. Chapter 10 in \cite{Saad:2003:IMS:829576}. The quality of the ILU decomposition can be tuned with the fill-in factor $k$, for which we test the values $k = 1, 2, 3$. Table \ref{tab:costs} (see also Fig. \ref{fig:linear_algebra_cost}) shows the cost of building the preconditioning matrix and the cost of GMRES iterations with the preconditioning operation, which both increase with $k$. Figure \ref{fig:preconditioner_giant} shows how ILU preconditioning affects the convergence of the GMRES method. For the red giant models, the figure shows that up to CFL$_\mathrm{hydro} \sim 40$, all ILU preconditioners have the same effect: they keep the number of GMRES iterations between 10 and 20. This is a net gain over the unpreconditioned GMRES that needs $\sim 500$ iterations to converge for the same time step. For large CFL numbers, the performances of ILU preconditioning clearly deteriorate. Note that ILU(3) does not show any improvement upon ILU(2). The results for the young Sun models show similar behavior. The major difference is that for CFL$_\mathrm{adv} \sim 1$, the number of iterations needed for convergence is roughly 200, i.e. significantly high. This suggests that a better preconditioner is needed to perform large CFL number computations, i.e. at CFL$_\mathrm{hydro} \gtrsim 1000$.

Finally, the cost of preconditioning can be amortized by recycling the preconditioning matrix during the computation. Indeed, since the preconditioner is only meant to be an approximation of $J^{-1}$, we find that it can be reused during Newton-Raphson iterations without losing much of its efficiency. Therefore, we compute the preconditioning matrix only at the first Newton-Raphson, and reuse it during the remaining Newton-Raphson iterations.




%
%
\section{Explicit methods}
\label{other_methods}

For comparison purpose, we also consider two explicit time-stepping schemes, which have a stability limit on the time step. The first scheme is a simple second-order explicit Adam-Bashforth method. The second scheme is the minimum residual approximate implicit scheme from \cite{botchev_stability_1999}, which we formulate as a stabilization of the Adam-Bashforth method.

\subsection{The explicit Adam-Bashforth method}


We use the second-order, explicit Adam-Bashforth method to discretize Eq. (\ref{eq:mol}), see e.g. Sect. 5.9 in \cite{leveque2007finite}. This method is a linear multistep method, and it requires one evaluation of the right-hand side $R$ per time step and the storage of the previous value $R(U^{n-1} )$. The update formula is

\begin{equation}
U^{n+1} = U^n + \frac{3}{2} \Delta t R(U^n) - \frac{1}{2} \Delta t R(U^{n-1} ).
\end{equation}

Because this method is explicit in time, it is prone to a CFL condition on the time step. We find that in our case a value of the CFL number as low as $\sim 0.1$ has to be used to perform a stable computation.

\subsection{The MRAI method}


The MRAI method is described in \cite{botchev_stability_1999}. A predictor step is performed with an explicit method, here the Adam-Bashforth method, with a larger time step than allowed for stability. A corrector step stabilizes the scheme by performing a \emph{fixed} amount of GMRES iterations (here taken to be five) using the implicit nonlinear residual (\ref{eq:crank_nicholson}). It can be seen as an explicit scheme with an update formula that adapts at each time step, thanks to the different action of the GMRES, to improve stability. It is still conditionally stable and hence limited to moderate values of the time step. In \cite{botchev_stability_1999}, an adaptive time step strategy is based on the highest eigenvalue of the Jacobian matrix. Here, we find that limiting the CFL number to a value of 1.5 was enough to maintain stability. This is 15 times more than the Adam-Bashforth method.

The advantage is that for such CFL numbers the use of preconditioning in the GMRES solver is not necessary (see Fig. \ref{fig:preconditioner_giant}). This allows taking advantage of an important property of Krylov methods: they only require the action of the Jacobian matrix on a vector, and not the Jacobian matrix itself. The action of the Jacobian can then be approximated by a finite difference:

\begin{equation}
J(\vec u) \vec v \approx \frac{F(\vec u + \delta \vec v) - F(\vec u)}{\delta},
\end{equation}

\noindent where $\delta$ is a small perturbation. The method is said to be ``Jacobian-free'', since the Jacobian matrix is never explicitly formed. The complete algorithm is

\begin{enumerate}
	\item Use the Adam-Bashforth method to perform a full time step, let $U^\star$ be the result;
	\item Perform five GMRES iterations on the system $J(U^\star) \delta U = - F(U^\star)$;
	\item Set $U^{n+1} = U^n + \delta U$;
	\item Limit the time step such that CFL $\lesssim 1.5$.
\end{enumerate}

In a Jacobian-free method, one GMRES iteration requires one nonlinear residual evaluation. Therefore, one time step of the MRAI solver requires six nonlinear residual evaluations: one to compute $F(U^\star)$, and five for the GMRES iterations.

%
%
\section{Results}
\label{results}

In this section, we first characterize the accuracy of the different nonlinear solvers, and then characterize the performances of the methods.

\subsection{Accuracy}
\label{accuracy}

In this section we characterize the accuracy of the implicit solvers based on the vortex advection test problem described in Sect. \ref{test_advection}.


\begin{table*}[t]
\caption{$L_1$-norm of the error for the vortex advection problem for a resolution of $256^2$, $\epsilon$ is the nonlinear tolerance of the Newton-Raphson procedure, and $\eta$ the linear tolerance of the Krylov solver.}
\centering
\begin{tabular}{l c c c}
\hline \hline
Method & $\epsilon = 10^{-2}$ & $\epsilon = 10^{-4}$ & $\epsilon = 10^{-6}$\\
\hline
Exact Newton                                  & 5.2710(-6)     & 5.3230(-7)	& 5.2788(-7) \\
\hline
Inexact Newton  ($\eta=10^{-6}$) & 5.2704(-6)     & 5.3226(-7)	& 5.2788(-7) \\
\ \ \ \ w/o preconditioner     & 5.2710(-6)     & 5.3233(-7)	& 5.2788(-7) \\
Inexact Broyden ($\eta=10^{-6}$) & 5.2704(-6)     & 5.6946(-7)	& 5.2788(-7) \\
\ \ \ \ w/o preconditioner     & 5.2710(-6)     & 5.6970(-7)	& 5.2788(-7) \\
\hline
Inexact Newton  ($\eta=10^{-4}$) & 5.2434(-6) & 5.3212(-7) & 5.2788(-7)\\
\ \ \ \ w/o preconditioner     &  5.4118(-6) &  5.3293(-7) &  5.2788(-7) \\
Inexact Broyden ($\eta=10^{-4}$) & 5.2434(-6)  & 5.6980(-7) & 5.2786(-7) \\
\ \ \ \ w/o preconditioner     &  1.1380(-5)& 5.7081(-7) &  5.2788(-7) \\
\hline
Inexact Newton  ($\eta=10^{-2}$)  & 6.6104(-6)     & 5.3091(-7)	& 5.2789(-7) \\
\ \ \ \ w/o preconditioner     & 1.1381(-5)     & 6.3107(-7)	& 5.2788(-7) \\
Inexact Broyden ($\eta=10^{-2}$)  & 6.6104(-6)     & 5.6907(-7)	& 5.2790(-7) \\
\ \ \ \ w/o preconditioner     & 1.1380(-5)     & 5.4528(-7)	& 5.2799(-7) \\
\hline
Inexact Newton  ($\eta=10^{-1}$)  & 3.6799(-5)     & 1.3171(-6)	& 5.2787(-7) \\
\ \ \ \ w/o preconditioner     & 8.1637(-5)     & 9.4765(-7)	& 5.2794(-7) \\
Inexact Broyden ($\eta=10^{-1}$)  & 3.6799(-5)     & 2.3078(-6)	& 5.2799(-7) \\
\ \ \ \ w/o preconditioner     & 8.1641(-5)     & 1.1158(-6)	& 5.2793(-7) \\
\hline
\end{tabular}
   \label{tab:norms256}
\end{table*}


%
\subsubsection{Sources of numerical errors}

We solve an initial value problem defined by a set of partial differential equations (Eqs (1-3)), appropriate boundary conditions, and some specified initial conditions. Let $u(x,t)$\footnote{For the sake of clarity, and without losing generality, we only refer to the $x$ direction in our notation.} be the exact solution of this problem. Let $U^n_i$ be the numerical solution of the discretized problem, and it is an approximation of $u(x_i, t^n)$\footnote{We consider here the pointwise value $u(x_i, t^n$), although we should refer to the volume-averaged quantity since we use finite volumes. This does not change our argument, and makes the notation lighter.}. In our case, $U_i^n$ is the solution of the discrete equations resulting from the Crank-Nicholson method and finite volumes discretization. Let $\mathcal{H}_{\Delta t}$ be the formal operator that corresponds to the \emph{exact} resolution of the nonlinear system (\ref{eq:crank_nicholson}):

\begin{equation}
\label{eq:H}
U^{n+1} = \mathcal{H}_{\Delta t} U^n  \Leftrightarrow F(U^{n+1})=0.
\end{equation}

\noindent However, the system of nonlinear equations is not solved exactly, and the Newton-Raphson solver introduces an error. Therefore, the numerical solution instead satisfies

\begin{equation}
\label{eq:H2}
U^{n+1}_i= \mathcal{H}_{\Delta t} U^n_i + \mathcal{E}^n_i,
\end{equation}

\noindent with $\mathcal{E}^n_i$ the error introduced by the Newton-Raphson solver. The local truncation error of the numerical scheme is defined as

\begin{align}
\label{eq:T}
T(x,t) &= \frac{1}{\Delta t} \big ( u(x, t+\Delta t) - \mathcal{H}_{\Delta t} u(x,t) \big ).
\end{align}

\noindent This corresponds to the error due to the discretization of the original partial differential equations. For the numerical scheme to be consistent, one requires that $T(x,t) \rightarrow 0$ as $\Delta x \rightarrow 0, \Delta t \rightarrow 0$. If the temporal discretization is of order $p$ and the spatial discretization of order $q$, then 

\begin{equation}
T(x,t) = \mathcal{O}(\Delta t^p) + \mathcal{O}(\Delta x^q).
\end{equation}

\noindent Depending on the ratio $\Delta t / \Delta x$, the truncation error can be either dominated by the temporal error or by the spatial error \citep[see for instance results and discussion in Appendix A of][]{viallet_towards_2011}.

The pointwise error  is defined as

\begin{align}
E^n_i = U^n_i - u(x_i, t^n).
\end{align}

\noindent From Eqs. (\ref{eq:H2}) and (\ref{eq:T}), we obtain

\begin{equation}
\label{eq:E}
E^{n+1}_i = \mathcal{H}_{\Delta t} E^n_i - \Delta t T(x_i,t^n) + \mathcal{E}^n_i.
\end{equation}

\noindent This equation illustrates how the new error stem from the old error (cumulative effect), and how the error created at each time step stem from the truncation error of the scheme and the error due to the inaccuracy in resolving the discrete equations. It is clear that it is desirable for $\mathcal{E}$ to be smaller than the truncation error.

The truncation error is ``controlled'' by the user, since it depends on the choice of the temporal and spatial discretization methods and on the spatial/temporal resolutions. As emphasized in the introduction, this paper does not address a comparison of different temporal methods and how their truncation errors affect the accuracy. However, an important issue that we do address here concerns the amplitude of $\mathcal{E}$. As shown by Eq. (\ref{eq:E}), it will affect the results when it dominates the truncation error, a very undesirable situation. What are the possible contributions to $\mathcal{E}$? First, note that for an explicit scheme, $\mathcal{E}$ is near the machine accuracy, so that in practice it is likely to be significantly smaller than truncation errors. For an implicit scheme, $\mathcal{E}$ can in principle be related to 1) the stopping criterion of the Newton-Raphson procedure, 2) the details of the nonlinear method itself (use of iterative methods, linear tolerance $\eta$, type of preconditioner, approximate Jacobian matrix, etc). The next section aims at elucidating this.

\subsubsection{Effects of nonlinear tolerance $\epsilon$ and nonlinear methods on accuracy}

\begin{figure*}[t] 
   \centering
   \parbox{0.32\linewidth}{\center \includegraphics[width=\linewidth]{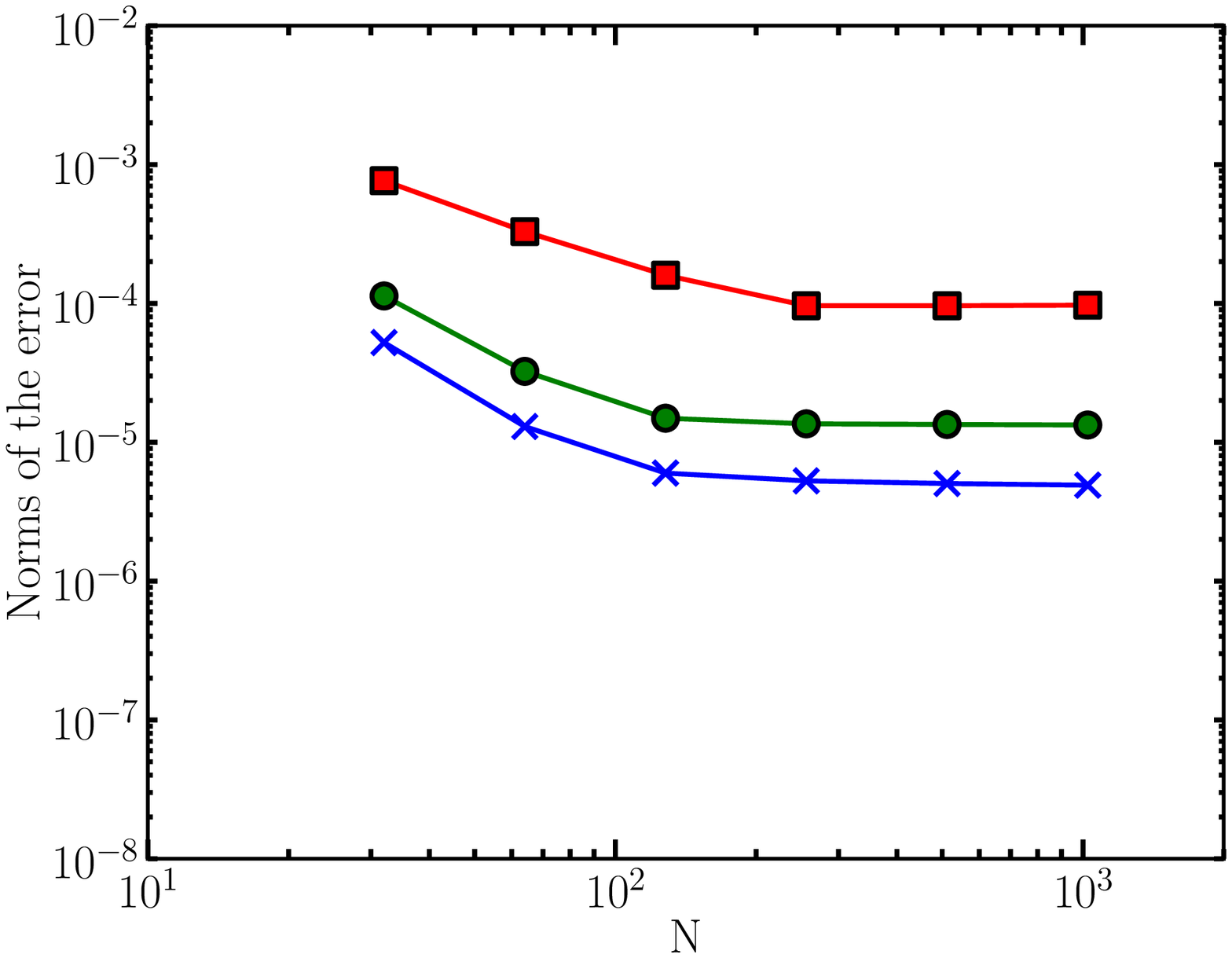}}
   \parbox{0.32\linewidth}{\center \includegraphics[width=\linewidth]{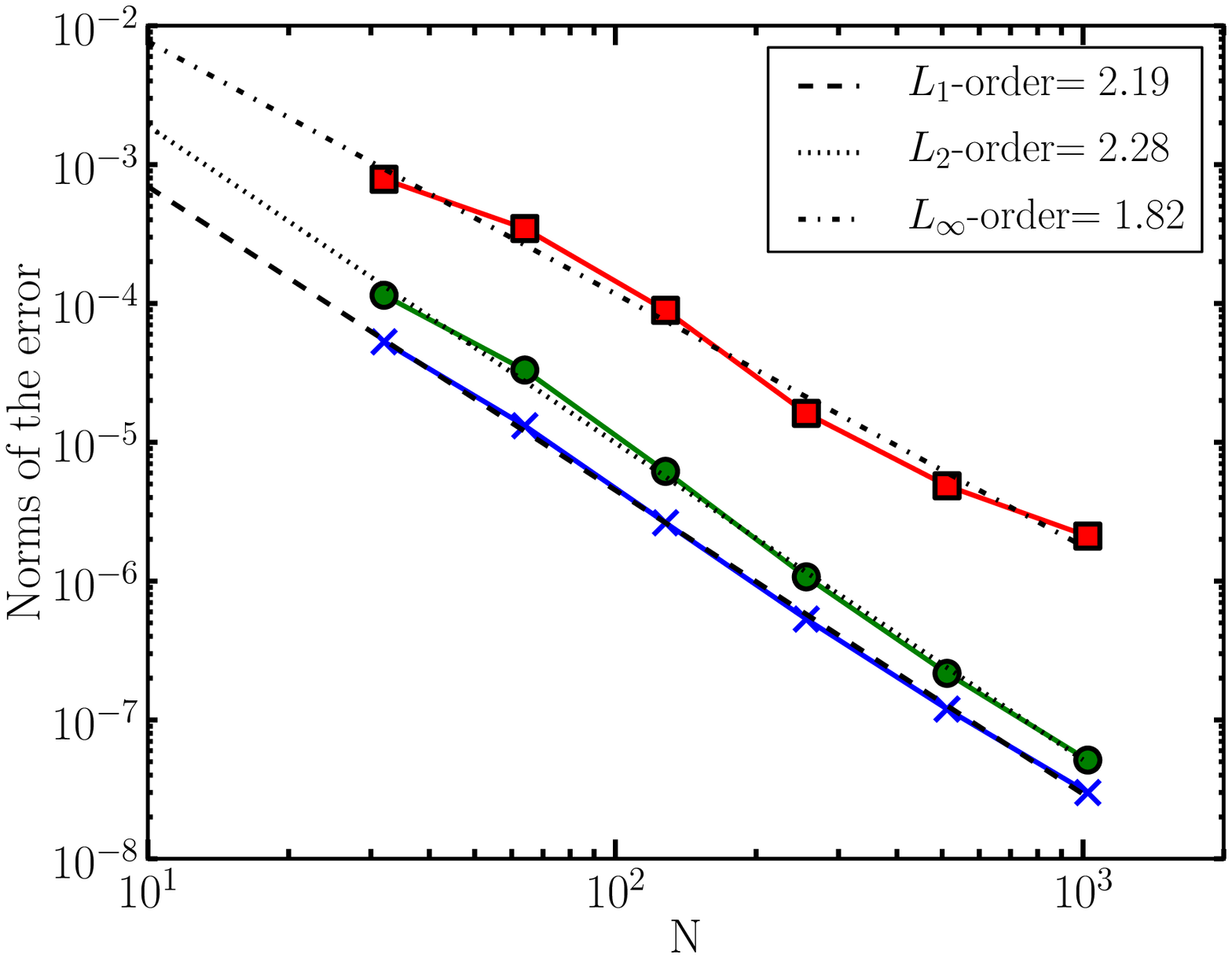}}
   \parbox{0.32\linewidth}{\center \includegraphics[width=\linewidth]{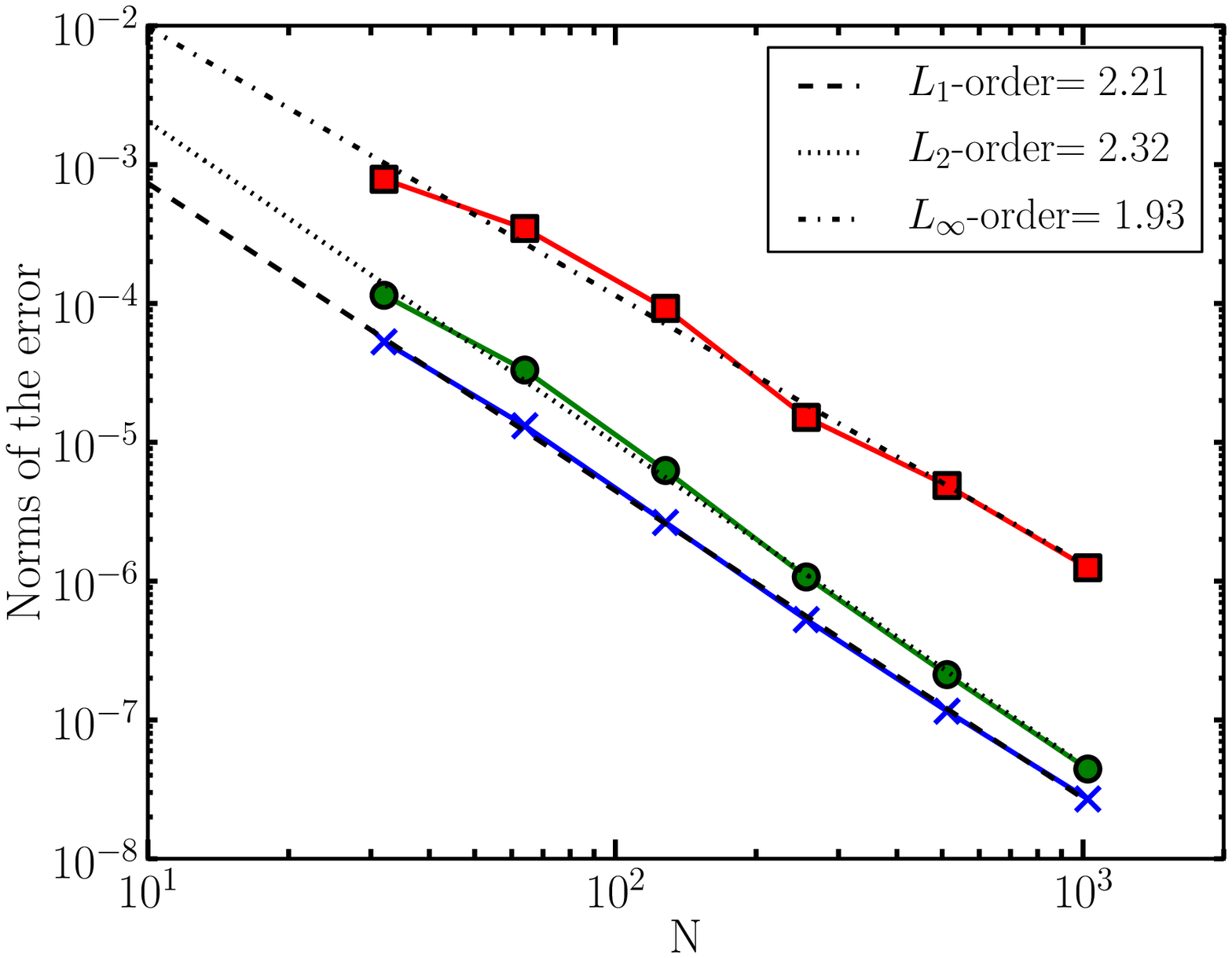}}
   \caption{Convergence of the numerical error with resolution for the vortex advection test. The error is measured with the $L_1$-norm (crosses), $L_2$-norm (dots), and $L_\infty$-norm (square). Results are obtained with the inexact Newton method ($\eta=10^{-6}$). Three different values of the nonlinear tolerance are investigated: $10^{-2}, 10^{-4}, 10^{-6}$ (from left to right).}
   \label{fig:newton_nl}
\end{figure*}

To determine the parameters that affect $\mathcal{E}$, we performed a large number of runs of the isentropic vortex advection test (see Sect. \ref{test_advection}) and monitored the numerical error. We used a sequence of uniform grids with resolutions ranging from $32^2$ to $1024^2$. Computations were done at \emph{constant} CFL$_\mathrm{hydro}$. We considered the same time step as \cite{kifonidis_multigrid_2012}, corresponding to CFL$_\mathrm{hydro} \sim 1.85$ and CFL$_\mathrm{adv} = 0.8$. With the ratio of $\Delta t / \Delta x$ fixed, the truncation error behaves as $\Delta x^2$ (or equivalently $\Delta t^2$), since our scheme is second order in space and time. In Sect. \ref{accuracy_dt}, we investigate the effect of the time step choice, keeping spatial resolution constant, on the accuracy.

The parameters we vary are

\begin{itemize}
\item the nonlinear tolerance $\epsilon$ in the stopping criterion of the Newton-Raphson iterations (see Table \ref{tab:generalNR}). We consider $\epsilon=10^{-2}$, $10^{-4}$, $10^{-6}$;
\item linear solver: ``exact'' method (direct solver MUMPS), ``inexact'' method (GMRES) with linear tolerance $\eta=10^{-1}$, $10^{-2}$, $10^{-4}$, $10^{-6}$;
\item Jacobian matrix strategy: Newton\footnote{i.e. the Jacobian matrix is recomputed at each Newton iteration.}, Broyden method;
\item preconditioner (iterative methods only): recycled ILU(1) or no preconditioner.
\end{itemize}

It should be emphasized that we could not use the direct solver MUMPS to perform test runs at resolutions $512^2$ and $1024^2$, because the memory requirement of the LU decomposition exceeded the available memory on the computer used to run the tests (32 gigabytes).

We first compare the results for all runs at a resolution of $256^2$. The error is measured by the difference between the computed density field and the analytical solution. The values of the numerical error, measured using the $L_1$-norm, are summarized in Table \ref{tab:norms256}. The ``Exact Newton'' with $\epsilon = 10^{-6}$ is the method used in \cite{viallet_towards_2011}. From the results shown in this table, we can draw important conclusions.

\begin{enumerate}
\item For a nonlinear tolerance $\epsilon = 10^{-6}$, \emph{all} methods show roughly the same numerical error (with three significant digits). In this case, the numerical accuracy does not depend on the details of the nonlinear method. For less strict nonlinear tolerances, however, we find that accuracy deteriorates for some nonlinear methods. In the worst cases, an order of magnitude can be lost in accuracy. We find that such cases often correspond to linear/nonlinear tolerances being lose enough so that only one (or possibly two) linear/nonlinear iterations are needed to reach convergence. Such situations should be avoided.
\item The inexact Newton method with linear tolerance $\eta=10^{-6}$ has roughly the same numerical error (with three significant digits) as the exact Newton method. This shows that the GMRES method with a strict tolerance yields similar results to the direct method MUMPS, which is much more expensive. The small difference in the error observed between the inexact Broyden method and the inexact Newton method, both with $\eta=10^{-6}$, can be attributed to the effect of the approximation of the Jacobian rather than to the inaccuracy of the linear solver. However, as noted above, the difference between both methods disappears when the nonlinear tolerance is strict enough.
\end{enumerate}

We now consider the inexact Newton method with $\eta=10^{-6}$, identified above as the most accurate method because it gives similar results to the direct solver MUMPS, independently of $\epsilon$. The memory requirement for iterative solvers are much less demanding, so we can use this method to study the convergence of the error with resolution for different nonlinear tolerance $\epsilon$. The results are shown in Fig. \ref{fig:newton_nl}. For $\epsilon=10^{-2}$, we find that the error saturates for resolutions higher than $64^2$. In this case, the error introduced by the loose nonlinear tolerance of the Newton iterations dominate the truncation error at resolutions higher than $64^2$. Other nonlinear methods lead, at best, to the same result. For nonlinear tolerances of $10^{-4}$ and $10^{-6}$, we find the expected second-order rate of convergence. In these cases, we conclude that $\mathcal{E}$ is smaller than the truncation error. From the results shown in Table \ref{tab:norms256}, it can be expected that at $\epsilon=10^{-4}$ some methods will \emph{not} exhibit a similar convergence, and that at $\epsilon=10^{-6}$ \emph{all} methods will exhibit the same convergence. We checked that this is indeed the case. We can therefore draw the important conclusion that, at least up to a resolution of $1024^2$, a nonlinear tolerance of $10^{-6}$ is strict enough to ensure that the numerical error is dominated by the truncation error stemming from the second-order temporal and spatial discretization.

This highlights that, in general, the nonlinear tolerance has to be tuned to match the expected value of the truncation error. As emphasized earlier, this depends on both the temporal/spatial discretizations (more precisely: their order of accuracy) and on the spatial/temporal resolutions. In principle, the truncation error can be made as small as desired (within the limit of machine accuracy). This means that a given nonlinear tolerance $\epsilon$ can limit the merits of high-resolution/high-order computations.

\subsubsection{Impact of the time step on accuracy}
\label{accuracy_dt}

In this section, we use the isentropic advection test problem to study how the numerical error depends on the time step. As a measure of the time step, we use the advective CFL number defined as

\begin{equation}
\mathrm{CFL}_\mathrm{adv} = \frac{u_\infty \Delta t}{\Delta x}.
\end{equation}

\noindent In the spirit of Eq. (\ref{CFLadv}), it measures by how many cells the vortex is advected across the grid during a time step. \cite{viallet_towards_2011} argued that $\mathrm{CFL}_\mathrm{adv} \gtrsim 1$ results in inaccurate advection, because the vortex is moved over several cells during a time step. To quantify this effect, we consider a constant resolution ($256^2$) and vary the time step in order to cover the broadest possible range of CFL$_\mathrm{adv}$. The maximum value we can reach is limited by the Newton-Raphson procedure: for too high values of CFL$_\mathrm{adv}$, the Newton-Raphson method does not converge any longer.

\begin{figure*}[t] 
   \centering
   \parbox{0.45\linewidth}{\center \includegraphics[width=0.8\linewidth]{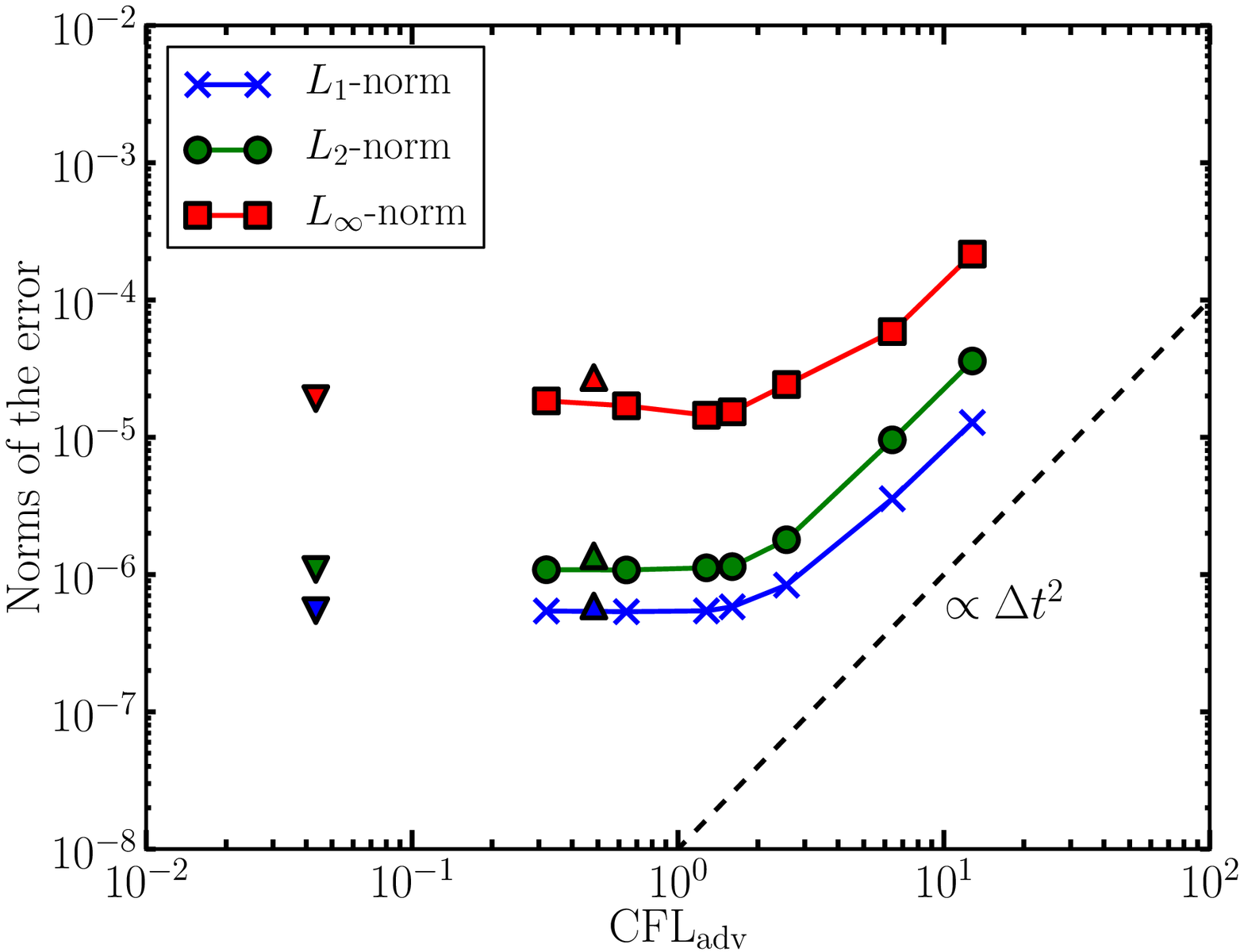}}
   \parbox{0.45\linewidth}{\center \includegraphics[width=0.8\linewidth]{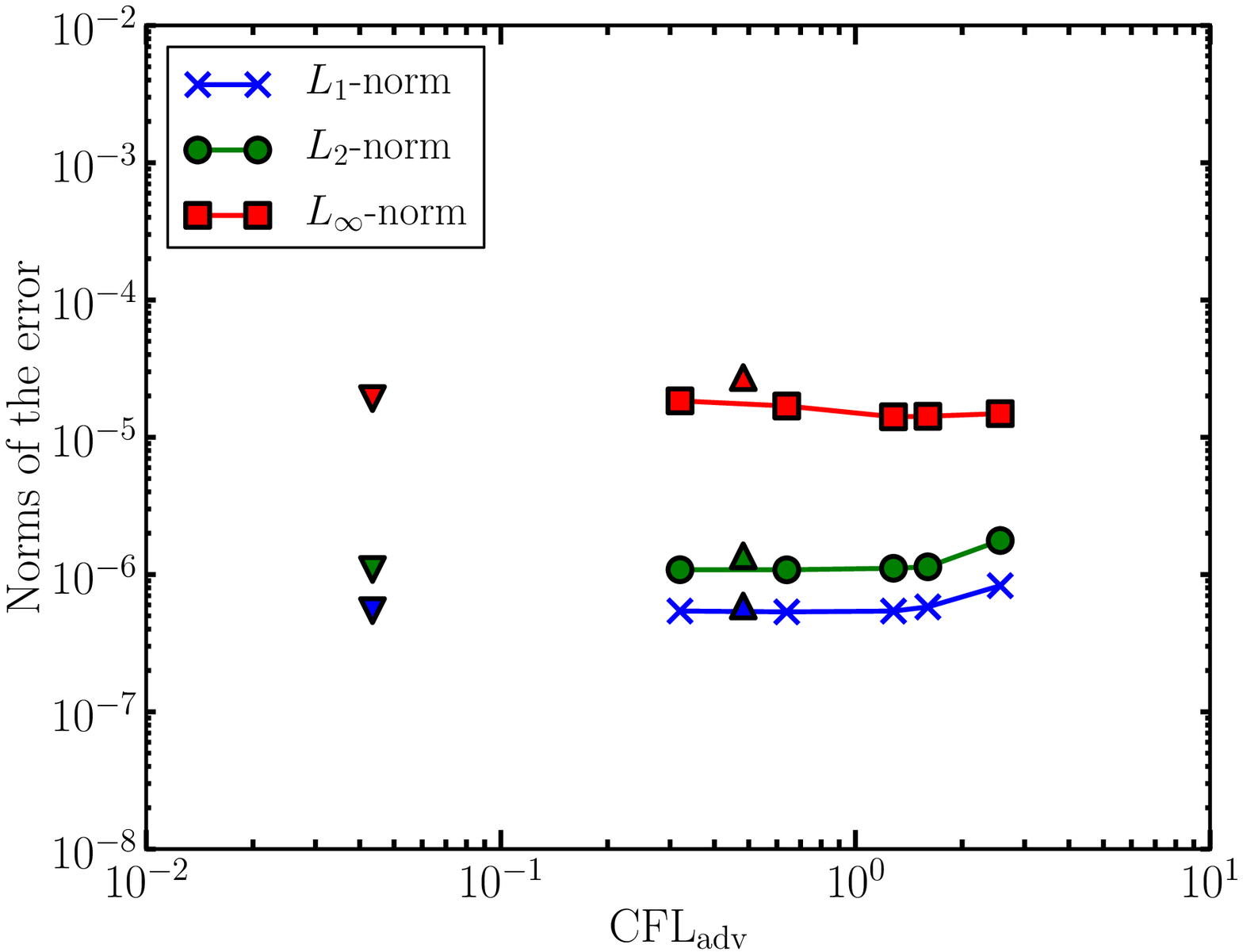}}
   \caption{Impact of the time step on the numerical accuracy of the vortex advection problem ($256^2$). The results were obtained with the inexact Newton method ($\eta=10^{-6}$), with $\epsilon=10^{-4}$ (left) and $\epsilon=10^{-6}$ (right). For comparison, the results obtained with the Adam-Bashforth (triangles down) and the MRAI (triangles up) methods are shown in both panels. The labels show the inferred rates of convergence of the different errors.}
   \label{fig:advection}
\end{figure*}

The test is performed with the inexact Newton method with $\eta=10^{-6}$. We consider $\epsilon=10^{-4}$ and $\epsilon=10^{-6}$, to ensure that the numerical error is dominated by the truncation error (see previous section). For $\epsilon=10^{-4}$, we can reach  CFL$_\mathrm{adv} \sim 10$, whereas for $\epsilon=10^{-6}$, the Newton-Raphson method does not converge when CFL$_\mathrm{adv} \gtrsim 3$. The results are presented in Fig. \ref{fig:advection}. These results show that when increasing the time step there is a transition from a regime where the numerical error is dominated by the spatial error (for CFL$_\mathrm{adv} \lesssim 1$), which is constant since the resolution is fixed, to a regime where the error is dominated by the temporal error  (CFL$_\mathrm{adv} \gtrsim 1$) and behaves as $\Delta t^2$. For $\epsilon=10^{-6}$, convergence difficulties act as a ``natural limitation'' of the time step and effectively prevent the numerical accuracy to deteriorate too significantly. The results obtained with the Adam-Bashforth and MRAI methods are indicated in Fig. \ref{fig:advection}. They show that our implicit methods are as accurate as these explicit methods when CFL$_\mathrm{adv} \lesssim 2$. It is worth noting that the $L_\infty$ error of the MRAI scheme is larger than the fully implicit scheme at a similar time step, whereas the $L_1$ and $L_2$ norms show that the global convergence is comparable to other schemes. This is due to the fixed number of Krylov iterations in the method, which does not ensure an uniform convergence over the computational domain. This shows that the MRAI scheme sacrifices somewhat pointwise accuracy for improved stability.

It should be emphasized that the situation investigated here is very idealized, because the test problem consists of a single vortex being advected by a uniform flow. More realistic flows, such as turbulent flows in stellar interiors, are instead characterized by a large number of eddies interacting in a chaotic way. The study of accuracy in such chaotic flow is complex so deserves further attention. Based on the results presented in this section, we consider that CFL$_\mathrm{adv} \lesssim 2$ is an acceptable condition on the time step to ensure accurate advection.

\subsection{Performances}

In the previous section, we showed that a nonlinear tolerance $\epsilon=10^{-6}$ ensures that the accuracy of the computation is not affected by the details of the nonlinear method. We now set $\epsilon=10^{-6}$ and benchmark the performances of the different nonlinear methods based on the stellar models presented in Sect. \ref{test_stellar_model}. The benchmark procedure is the following. We start our runs from 2D initial models that provide a state where convective motions are already well-developed and in a quasi-steady state (i.e., the models were already advanced for some time prior to the benchmark). These initial snapshots are shown in Fig. \ref{fig:initial_models}. We perform 100 time steps using the different methods described in Sect. \ref{fullyimplicit}. The strategy on the time step is based on a limitation of the advective CFL, with $\mathrm{CFL}_\mathrm{adv,max} = 0.5,\ 1,\ 1.5,\ 2$. We do not consider larger time steps, for which accuracy deteriorates significantly (see previous section). For the two explicit methods described in Sect. \ref{other_methods}, we choose the time step according to stability consideration. As the resulting time step is significantly smaller, we compute more models, typically 1000 for the MRAI method, and 10 000 for the Adam-Bashforth method.

We performed a large set of test runs to study how the performances are affected by 1) the choice between the Newton and the Broyden methods, 2) the tolerance $\eta$ on the linear solver, 3) the fill-in level of the ILU preconditioner. We also discuss how the choice of the time step affects the performances. The results are summarized in Table \ref{tab:rg_runs_summary} for the red giant models and Table \ref{tab:ys_runs_summary} for the young Sun models. For the sake of clarity, we do not show the results of all runs but only those which illustrate the main conclusions drawn in the discussion below.

Our results show that the most critical choice in obtaining good performances is to use the Broyden method rather than the Newton method. Although the average number of nonlinear iterations per time step is larger with the Broyden method, computing the Broyden update of the Jacobian matrix is much faster than recomputing the Jacobian matrix. Furthermore, our results show that the average number of nonlinear iterations per time step is not strongly affected by the value of the tolerance $\eta$ on the linear solver. Therefore, an inexact strategy with a low tolerance ($\eta = 10^{-1}, 10^{-2}$) leads to better performances than with a more strict tolerance (e.g. $\eta=10^{-4}$). The choice of ILU preconditioner has to be made according to the average value of CFL$_\mathrm{hydro}$. For the red giant models, which have rather low CFL$_\mathrm{hydro}$, ILU(1) performs the best. For the young Sun models, the ILU(2) preconditioner is the best choice: the CFL$_\mathrm{hydro}$ is too large for ILU(1) to be efficient (see Fig. \ref{fig:preconditioner_giant}) and we find that ILU(3) does not decrease the number of GMRES iterations significantly to compensate for its cost. This ``saturation" in efficiency of the preconditioner has already been discussed with Fig. \ref{fig:preconditioner_giant}. With a better preconditioner, it is not excluded that better performances could be achieved for the young Sun models. We come back to the shortcomings of our present preconditioning method in the conclusion. Finally, it is important to recycle the preconditioning matrix to obtain good performances, since building the ILU factorization has a cost similar to that of the Jacobian matrix computation. We tested that recycling the preconditioner during the whole Newton-Raphson procedure has no impact on the number of nonlinear iterations, thus leading to the best performances.

Concerning the time step, the efficiency of the solvers increases with the values of CFL$_\mathrm{adv}$. The average number of Newton-Raphson iterations per time step increases with the advective CFL number, since the solution varies more between two time steps, achieving convergence with more Newton-Raphson iterations \citep[see discussion in][]{viallet_towards_2011}. However, this increase in the number of nonlinear iterations is compensated for by the longer time step, leading to the observed gain in efficiency. This shows that an aggressive strategy on the time step is rewarded by the best performances.

Finally, a comparison with the performance of explicit methods shows that our best implicit solvers perform faster, leading to speed-up factors of roughly $2-3$ for the red giant calculations and roughly $15-20$ for the young Sun calculations. Nevertheless, it should be emphasized that our explicit methods are stable only for quite low values of the time step, but the \MUSIC\ code was not optimized for explicit calculations. For instance, the second-order Adam-Bashforth is computational inexpensive, but it has a very strict stability limit. In the future, we plan to benchmark the performances of the \MUSIC\ code against other, ``state-of-the-art", explicit codes. 

\begin{table*}[t]
   \caption{Summary of the tests for the red giant. Nomenclature for the methods. ``Broyden($10^{-2}$)'': Broyden method with a tolerance $\eta=10^{-2}$ on the linear solver. ``rILU(k)": recycled ILU preconditioner with fill-in factor $k$. The columns: the mean values of the three CFL numbers defined in Sect. \ref{testcases}, the average number of Newton-Raphson iterations per time step, the average number of GMRES iterations per Newton-Raphson iteration, the percentage of the wall time spent in: i) computing the Jacobian matrix, ii) building the ILU preconditioner, iii) performing the GMRES solve, and finally the ratio of the simulated time by the wall time, which measures the computational efficiency of the method. Methods are shown in each subsection in order of decreasing efficiency. In each case, only the most efficient Newton method is shown.}
   \label{tab:rg_runs_summary}
   \centering
  
\begin{tabular}{l c c c c c c c c c}
\hline \hline
Method & CFL$_\mathrm{hydro}$ &  CFL$_\mathrm{adv}$ & CFL$_\mathrm{rad}$ & $\frac{\mathrm{Newton}}{\Delta t}$ & $\frac{\mathrm{GMRES}}{\mathrm{Newton}}$ & $\frac{\mathrm{Jacobian\ time}}{\mathrm{Wall\ time}}$ & $\frac{\mathrm{ILU\ time}}{\mathrm{Wall\ time}}$ & $\frac{\mathrm{GMRES\ time}}{\mathrm{Wall\ time}}$ & $\frac{\mathrm{Simulated\ time}}{\mathrm{Wall\ time}}$ \\
{\bf Explicit schemes}\\ 
MRAI & 1.50 & $3.8(-2)$ & $0.28$ & - & - & - & - & - & 371\\
Adam-Bashforth & 0.10 & $2.5(-3)$ & $1.5(-2)$ & - & - & - & - & - & 201\\
\hline
{\bf Implicit schemes}\\
{\bf CFL$_\mathrm{adv,max} = 0.5$} \\
Broyden($10^{-2}$) - rILU(1) & 19.1 & 0.50 & 5.0 & 7.3 & 4.2 & 56.7\% & 13.9\% & 11.4\% & 404\\
Broyden($10^{-1}$) - rILU(1) & 19.1 & 0.50 & 5.0 & 7.7 & 2.8 & 57.7\% & 13.5\% & 10.1\% & 390\\
Broyden($10^{-4}$) - rILU(1) & 19.1 & 0.50 & 5.0 & 7.3 & 7.2 & 53.3\% & 12.5\% & 17.5\% & 364\\
Broyden($10^{-1}$) - rILU(2) & 19.1 & 0.50 & 5.0 & 7.7 & 2.5 & 47.5\% & 26.9\% & 9.7\% & 336\\
Broyden($10^{-2}$) - rILU(2) & 19.1 & 0.50 & 5.0 & 7.3 & 3.8 & 46.3\% & 26.4\% & 12.3\% & 330\\
Broyden($10^{-4}$) - rILU(2) & 19.1 & 0.50 & 5.0 & 7.3 & 6.8 & 42.6\% & 23.1\% & 20.8\% & 287\\
Broyden($10^{-2}$) - rILU(3) & 19.1 & 0.50 & 5.0 & 7.3 & 3.7 & 33.2\% & 41.3\% & 14.5\% & 237\\
Broyden($10^{-1}$) - rILU(3) & 19.1 & 0.50 & 5.0 & 7.7 & 2.5 & 34.6\% & 41.2\% & 12.4\% & 235\\
Broyden($10^{-4}$) - rILU(3) & 19.1 & 0.50 & 5.0 & 7.3 & 6.7 & 31.1\% & 38.5\% & 20.2\% & 221\\
Newton($10^{-2}$) - rILU(1) & 19.1 & 0.50 & 5.0 & 5.2 & 4.5 & 87.5\% & 4.9\% & 3.0\% & 140\\
\hline
{\bf CFL$_\mathrm{adv,max} = 1$} \\
Broyden($10^{-1}$) - rILU(1) & 40.2 & 0.99 & 8.7 & 9.1 & 4.5 & 53.4\% & 12.7\% & 13.8\% & 779\\
Broyden($10^{-2}$) - rILU(1) & 40.2 & 0.99 & 8.7 & 8.7 & 6.5 & 50.7\% & 11.8\% & 19.2\% & 719\\
Broyden($10^{-1}$) - rILU(2) & 40.2 & 0.98 & 8.6 & 9.3 & 3.5 & 44.4\% & 24.4\% & 13.9\% & 659\\
Broyden($10^{-2}$) - rILU(2) & 40.4 & 0.99 & 8.8 & 8.7 & 5.4 & 42.8\% & 23.8\% & 17.6\% & 632\\
Broyden($10^{-4}$) - rILU(1) & 40.4 & 0.99 & 8.8 & 8.7 & 11.7 & 44.1\% & 10.3\% & 29.7\% & 632\\
Broyden($10^{-4}$) - rILU(2) & 40.4 & 0.99 & 8.8 & 8.7 & 10.0 & 37.8\% & 21.1\% & 27.2\% & 559\\
Broyden($10^{-1}$) - rILU(3) & 40.4 & 0.99 & 8.7 & 9.1 & 3.4 & 31.9\% & 39.8\% & 15.8\% & 473\\
Broyden($10^{-2}$) - rILU(3) & 40.2 & 0.99 & 8.7 & 8.8 & 5.1 & 30.9\% & 38.7\% & 18.8\% & 454\\
Broyden($10^{-4}$) - rILU(3) & 40.4 & 0.99 & 8.8 & 8.7 & 9.7 & 26.6\% & 33.4\% & 30.1\% & 393\\
Newton($10^{-2}$) - rILU(1) & 40.5 & 0.99 & 8.8 & 5.9 & 7.3 & 86.8\% & 4.3\% & 4.5\% & 262\\
\hline
{\bf CFL$_\mathrm{adv,max} = 1.5$} \\
Broyden($10^{-1}$) - rILU(1) & 57.3 & 1.49 & 11.7 & 11.0 & 5.5 & 49.1\% & 11.3\% & 18.0\% & 1008\\
Broyden($10^{-2}$) - rILU(1) & 57.3 & 1.49 & 11.7 & 10.5 & 8.3 & 45.5\% & 10.1\% & 25.4\% & 896\\
Broyden($10^{-2}$) - rILU(2) & 57.3 & 1.49 & 11.7 & 10.6 & 6.5 & 39.3\% & 21.1\% & 22.7\% & 812\\
Broyden($10^{-1}$) - rILU(2) & 55.7 & 1.42 & 10.8 & 12.4 & 4.0 & 41.0\% & 21.6\% & 18.1\% & 767\\
Broyden($10^{-4}$) - rILU(1) & 57.3 & 1.49 & 11.7 & 10.5 & 19.6 & 33.4\% & 7.5\% & 45.2\% & 664\\
Broyden($10^{-2}$) - rILU(3) & 57.3 & 1.49 & 11.7 & 10.6 & 6.2 & 28.9\% & 34.0\% & 24.5\% & 599\\
Broyden($10^{-4}$) - rILU(2) & 57.3 & 1.49 & 11.7 & 10.5 & 17.6 & 28.5\% & 15.3\% & 44.0\% & 590\\
Broyden($10^{-1}$) - rILU(3) & 55.8 & 1.43 & 10.9 & 12.3 & 3.9 & 29.7\% & 36.0\% & 20.4\% & 541\\
Broyden($10^{-4}$) - rILU(3) & 57.3 & 1.49 & 11.7 & 10.5 & 17.2 & 18.6\% & 21.3\% & 52.2\% & 374\\
Newton($10^{-2}$) - rILU(1) & 56.2 & 1.45 & 11.1 & 7.5 & 8.8 & 86.8\% & 3.4\% & 5.5\% & 286\\
\hline
{\bf CFL$_\mathrm{adv,max} = 2$} \\
Broyden($10^{-1}$) - rILU(1) & 70.4 & 1.95 & 13.9 & 13.5 & 6.9 & 43.4\% & 9.3\% & 25.3\% & 1015\\
Broyden($10^{-2}$) - rILU(1) & 70.2 & 1.94 & 14.0 & 13.1 & 10.5 & 40.5\% & 9.0\% & 30.7\% & 960\\
Broyden($10^{-2}$) - rILU(2) & 69.7 & 1.92 & 13.9 & 13.6 & 7.5 & 35.7\% & 18.4\% & 28.0\% & 813\\
Broyden($10^{-1}$) - rILU(2) & 67.9 & 1.85 & 13.8 & 13.8 & 4.6 & 38.5\% & 18.6\% & 24.1\% & 785\\
Broyden($10^{-1}$) - rILU(3) & 70.4 & 1.95 & 13.9 & 13.8 & 4.5 & 29.0\% & 32.5\% & 23.3\% & 696\\
Broyden($10^{-2}$) - rILU(3) & 70.2 & 1.94 & 14.1 & 13.1 & 6.9 & 25.4\% & 27.7\% & 34.4\% & 576\\
Broyden($10^{-4}$) - rILU(1) & 69.2 & 1.90 & 14.0 & 13.4 & 50.3 & 16.3\% & 3.5\% & 72.4\% & 353\\
Broyden($10^{-4}$) - rILU(2) & 70.2 & 1.94 & 14.1 & 13.4 & 41.4 & 14.2\% & 7.2\% & 71.5\% & 329\\
Newton($10^{-1}$) - rILU(1) & 69.5 & 1.91 & 13.9 & 9.9 & 7.7 & 88.3\% & 2.6\% & 4.9\% & 269\\
Broyden($10^{-4}$) - rILU(3) & 68.6 & 1.89 & 13.8 & 13.4 & 60.5 & 7.4\% & 8.0\% & 81.1\% & 163\\
\end{tabular}
\end{table*}

\begin{table*}[t]
   \caption{Same as Table \ref{tab:rg_runs_summary}, but for the young Sun models.}
   \label{tab:ys_runs_summary}
   \centering
   
\begin{tabular}{l c c c c c c c c c}
\hline \hline
Method & CFL$_\mathrm{hydro}$ &  CFL$_\mathrm{adv}$ & CFL$_\mathrm{rad}$ & $\frac{\mathrm{Newton}}{\Delta t}$ & $\frac{\mathrm{GMRES}}{\mathrm{Newton}}$ & $\frac{\mathrm{Jacobian\ time}}{\mathrm{Wall\ time}}$ & $\frac{\mathrm{ILU\ time}}{\mathrm{Wall\ time}}$ & $\frac{\mathrm{GMRES\ time}}{\mathrm{Wall\ time}}$ & $\frac{\mathrm{Simulated\ time}}{\mathrm{Wall\ time}}$ \\
\hline
{\bf Explicit schemes}\\ 
MRAI & 1.50 & $5.9(-3)$ & $1.6(-9)$ & - & - & - & - & - & 13\\
Adam-Bashforth & 0.10 & $2.8(-4)$ & $10^{-10}$ & - & - & - & - & - & 7\\
\hline
{\bf Implicit schemes}\\
{\bf CFL$_\mathrm{adv,max} = 0.5$} \\
Broyden($10^{-1}$) - rILU(2) & 235.7 & 0.50 & 2.5(-7) & 6.6 & 12.8 & 38.1\% & 21.5\% & 28.5\% & 124\\
Broyden($10^{-2}$) - rILU(2) & 235.7 & 0.50 & 2.5(-7) & 5.9 & 20.5 & 33.9\% & 19.4\% & 37.1\% & 112\\
Broyden($10^{-1}$) - rILU(3) & 235.7 & 0.50 & 2.5(-7) & 6.6 & 11.8 & 28.2\% & 33.7\% & 29.1\% & 92\\
Broyden($10^{-4}$) - rILU(2) & 235.7 & 0.50 & 2.5(-7) & 5.9 & 33.9 & 26.7\% & 15.3\% & 50.4\% & 88\\
Broyden($10^{-2}$) - rILU(3) & 235.7 & 0.50 & 2.5(-7) & 5.9 & 19.1 & 24.2\% & 29.4\% & 39.4\% & 80\\
Broyden($10^{-4}$) - rILU(3) & 235.7 & 0.50 & 2.5(-7) & 5.9 & 31.3 & 18.6\% & 22.7\% & 53.2\% & 62\\
Newton($10^{-2}$) - rILU(2) & 235.7 & 0.50 & 2.5(-7) & 4.5 & 24.5 & 69.3\% & 9.6\% & 17.3\% & 56\\
\hline
{\bf CFL$_\mathrm{adv,max} = 1$} \\
Broyden($10^{-1}$) - rILU(2) & 474.0 & 1.00 & 5.1(-7) & 7.2 & 21.4 & 30.5\% & 17.1\% & 42.3\% & 199\\
Broyden($10^{-2}$) - rILU(2) & 474.0 & 1.00 & 5.1(-7) & 6.5 & 35.9 & 24.8\% & 14.1\% & 53.6\% & 163\\
Broyden($10^{-1}$) - rILU(3) & 474.0 & 1.00 & 5.1(-7) & 7.2 & 19.8 & 22.1\% & 26.1\% & 44.4\% & 143\\
Broyden($10^{-2}$) - rILU(3) & 474.0 & 1.00 & 5.1(-7) & 6.5 & 32.7 & 17.7\% & 21.3\% & 55.5\% & 117\\
Broyden($10^{-4}$) - rILU(2) & 474.0 & 1.00 & 5.1(-7) & 6.4 & 68.3 & 16.7\% & 9.5\% & 68.8\% & 110\\
Newton($10^{-2}$) - rILU(2) & 474.4 & 1.00 & 5.1(-7) & 4.6 & 44.9 & 60.3\% & 8.0\% & 28.6\% & 93\\
Broyden($10^{-4}$) - rILU(3) & 474.0 & 1.00 & 5.1(-7) & 6.4 & 63.0 & 12.5\% & 15.0\% & 68.8\% & 82\\
\hline
{\bf CFL$_\mathrm{adv,max} = 1.5$} \\
Broyden($10^{-1}$) - rILU(2) & 680.8 & 1.50 & 7.3(-7) & 8.1 & 29.5 & 24.6\% & 13.6\% & 53.0\% & 227\\
Broyden($10^{-2}$) - rILU(2) & 680.8 & 1.50 & 7.3(-7) & 7.4 & 48.4 & 19.3\% & 10.8\% & 63.5\% & 180\\
Broyden($10^{-1}$) - rILU(3) & 677.7 & 1.49 & 7.3(-7) & 8.0 & 27.0 & 18.1\% & 20.5\% & 55.1\% & 158\\
Broyden($10^{-2}$) - rILU(3) & 680.8 & 1.50 & 7.3(-7) & 7.5 & 44.2 & 13.7\% & 16.2\% & 65.3\% & 128\\
Newton($10^{-2}$) - rILU(2) & 681.0 & 1.50 & 7.3(-7) & 4.9 & 62.2 & 54.6\% & 7.1\% & 35.5\% & 118\\
Broyden($10^{-4}$) - rILU(2) & 680.8 & 1.50 & 7.3(-7) & 7.4 & 100.1 & 11.4\% & 6.3\% & 78.6\% & 106\\
Broyden($10^{-4}$) - rILU(3) & 680.8 & 1.50 & 7.3(-7) & 7.5 & 91.2 & 8.2\% & 9.7\% & 79.4\% & 76\\
\hline
{\bf CFL$_\mathrm{adv,max} = 2$} \\
Broyden($10^{-1}$) - rILU(2) & 914.6 & 2.00 & 9.8(-7) & 9.3 & 37.9 & 19.7\% & 10.7\% & 61.8\% & 239\\
Broyden($10^{-2}$) - rILU(2) & 914.6 & 2.00 & 9.8(-7) & 8.8 & 60.4 & 14.9\% & 8.1\% & 71.3\% & 182\\
Broyden($10^{-1}$) - rILU(3) & 914.6 & 2.00 & 9.8(-7) & 9.3 & 34.0 & 14.4\% & 16.5\% & 63.2\% & 175\\
Broyden($10^{-2}$) - rILU(3) & 915.0 & 2.00 & 9.9(-7) & 8.7 & 54.8 & 11.0\% & 12.5\% & 72.4\% & 132\\
Newton($10^{-1}$) - rILU(2) & 916.8 & 2.00 & 9.9(-7) & 6.8 & 43.2 & 63.0\% & 5.8\% & 28.1\% & 130\\
Broyden($10^{-4}$) - rILU(2) & 915.9 & 2.00 & 9.9(-7) & 8.8 & 142.0 & 7.6\% & 4.1\% & 85.5\% & 93\\
Broyden($10^{-4}$) - rILU(3) & 915.9 & 2.00 & 9.9(-7) & 8.7 & 128.7 & 5.8\% & 6.7\% & 85.4\% & 71\\
\end{tabular}
\end{table*}

\section{Conclusion}
\label{conclusion}


This paper presented a comparison of different Newton-Raphson solvers for time-implicit hydrodynamical computations. We first elucidated the importance of using a strict tolerance on the convergence criteria of the Newton-Raphson procedure: it ensures that the numerical error is dominated by truncation errors rather than by errors related to the nonlinear method itself. In our case, we found that requiring relative corrections  to become smaller than $\epsilon=10^{-6}$ is optimal given the typical grid resolutions we consider.

Having ascertained that the different nonlinear methods do not affect accuracy, we then benchmarked their performances. We identified the Broyden method as the most efficient nonlinear solver as compared to the standard Newton method. We show that an ``inexact'' strategy on the linear system leads to better performances. We identify the preconditioner as a crucial component of the method, since preconditioning is inherently required in the large CFL number regime. We used an incomplete LU factorization of the Jacobian matrix. The cost of building the ILU preconditioning matrix is important, but its impact on the performance of the method can be significantly mitigated by recycling the preconditioner. We find that the optimum recycling strategy is to build the ILU factorization only at the first Newton-Raphson iteration.

Concerning the choice of the time step, we showed that advection is resolved accurately when the time step fulfills CFL$_\mathrm{adv} \lesssim 2$. It should be stressed that this conclusion was drawn from a simplified physical situation, i.e. the advection of a single vortex by a uniform flow. For turbulent flows, a pointwise study of accuracy is undermined by the chaotic nature of the flow. One possibility is to study convergence of meaningful statistical properties of flow, as for instance the mean convective flux. We plan to investigate this in the future, but it can be expected that averaged quantities will be quite robust. Knowing the maximum optimal time step for accuracy is important, because our results show that the best performances are obtained when allowing for a large number of nonlinear iterations resulting from an ``aggressive" choice for the time step.

Our results show that an effective implicit solver, at least in 2D, can be built based on already existing software libraries, such as {\sf Trilinos}. These toolkits provide state-of-the-art implementations of the building blocks needed for an implicit solver: quasi-Newton methods, Krylov solvers, colored finite algorithm to compute the Jacobian matrix, black-box algebraic preconditioners, etc. The methods presented here perform significantly better than the explicit methods implemented in the \MUSIC\ code, but there is room for improvement. We identify two bottlenecks: the computation of the Jacobian matrix, and the ILU preconditioner, which becomes inefficient at large CFL numbers. For the first, the so-called ``Jacobian-free" approach is an attractive possibility. Jacobian-free Newton-Krylov solvers are very popular methods of solving large-scale problems; see \cite{knoll_jacobian-free_2004} for a review. The Jacobian-free GMRES method needs one evaluation of the nonlinear residual at each iteration. Using CFD, a Jacobian matrix in 2D is computed in roughly 50 nonlinear residual evaluations (see Sect. \ref{jacobian}), which means that a Jacobian-free GMRES becomes interesting if the total number of GMRES iterations can be kept below 50. This naturally leads us to consider the second bottleneck, which is the preconditioning method. 

We show here that a simple ILU strategy is only efficient at CFL$_\mathrm{hydro} \lesssim 100$. The ``black-box'' nature of ILU factorizations is an advantage both in terms of flexibility and ease of implementation, especially since it is widely implemented in existing scientific libraries. Other possible algebraic preconditioners are variants of incomplete factorization as ILU with threshold (ILUT) or the block incomplete LU preconditioner described in \cite{Ploeg_blockincomplete}; algebraic multigrid (AMG) preconditioners; sparse approximate inverse (SAI) preconditioners \citep[see e.g.][]{Wang2009409}, see e.g. \cite{Saad:2003:IMS:829576} for an introduction to ILUT, AMG, and SAI. It is important to stress that algebraic preconditioners are based on the coefficient matrix and completely ignore the physical processes responsible for the bad conditioning of the Jacobian matrix. On the other hand, physics-based preconditioners \citep[see e.g.][]{knoll_jacobian-free_2004} are based on the physical processes responsible for numerical stiffness, i.e. sound waves and/or thermal diffusion, rather than on the structure/elements of the matrix. The implementation of such a preconditioning method is left for a future work.

Finally, the analysis of the implicit solvers presented in this work begs an important question within the context of stellar hydrodynamics applications: Can we use them to perform 3D calculations? In 3D, the Jacobian matrix has a more complex sparsity pattern, with new extra diagonal terms due to the third dimension. As a result, we find that the number of colors of the Jacobian is multiplied by a factor of two in 3D, i.e. $n_g = 100$, which makes Jacobian-free methods even more attractive. Similarly, the incomplete LU factorization is more complex and more expensive to compute. The methods presented in this work would certainly work, but not efficiently. We expect to face the same issues regarding the inefficiency of ILU preconditioning at large CFL numbers. Our efforts are now devoted to the development of a better preconditioner and the use of Jacobian-free Krylov methods, with the aim of achieving efficient implicit computations at large CFL numbers both in 2D and 3D. Furthermore, with the perspective of performing large-scale 3D computations, we will address the challenging question of the parallelization and scalability of our implicit methods in future publications. 

\begin{acknowledgement}
MV acknowledges support  from a Newton International Fellowship and Alumni program from the Royal Society. Part of this work was funded by the Royal Society Wolfson Merit award WM090065, the Consolidated STFC grant  ST/J001627/1STFC and by the French ``Programme National de Physique Stellaire'' (PNPS). Finally, the authors thank an anonymous referee for his/her constructive criticism that lead to a substantial improvement of the paper.
\end{acknowledgement}

\bibliographystyle{aa}
\bibliography{references}

\end{document}